\documentclass[preprint,3p,12pt]{elsarticle}
\usepackage{amsmath}
\usepackage{stmaryrd}
\usepackage{bbding}
\usepackage{dcolumn}
\usepackage{graphicx}
\usepackage{amsfonts}
\usepackage{amssymb}
\usepackage{psfrag}
\usepackage{wrapfig}
\usepackage{subfigure}
\usepackage{makeidx}
\usepackage{bm}
\usepackage{epsf}
\usepackage{epsfig}
\usepackage{setspace}
\usepackage{color}
\usepackage{epstopdf}
\newtheorem{rmk}{Remark}

\begin{document}
\title{High-order gas-kinetic scheme with parallel computation for direct numerical simulation of turbulent flows}

\author[HKUST1]{Guiyu Cao}
\ead{gcaoaa@connect.ust.hk}

\author[BNU]{Liang Pan\corref{cor}}
\ead{panliang@bnu.edu.cn}

\author[HKUST1,HKUST2]{Kun Xu}
\ead{makxu@ust.hk}

\address[HKUST1]{Department of Mathematics, Hong Kong University of Science and Technology, Clear Water Bay, Kowloon, Hong Kong}
\address[BNU]{School of Mathematical Science, Beijing Normal University, Beijing, China}
\address[HKUST2]{Shenzhen Research Institute, Hong Kong University of Science and Technology, Shenzhen, China}
\cortext[cor]{Corresponding author}

\begin{abstract}
The performance of high-order gas-kinetic scheme (HGKS) has been
investigated for the direct numerical simulation (DNS) of isotropic
compressible turbulence up to the supersonic regime
\cite{GKS-high-4}. Due to the multi-scale nature
and coupled temporal-spatial evolution process,
HGKS provides a valid tool for the
numerical simulation of compressible turbulent flow.
Based on the domain decomposition and message passing interface (MPI), a parallel HGKS
code is developed for large-scale computation in this paper. The
standard tests from the nearly incompressible flow to the
supersonic one, including Taylor-Green vortex problem, turbulent channel
flow and isotropic compressible turbulence, are presented  to
validate  the parallel scalability, efficiency, accuracy and
robustness of parallel implementation. The performance of HGKS
for the nearly incompressible turbulence is comparable with the high-order
finite difference scheme, including the resolution of flow structure and efficiency of computation.
Based on the accuracy of the numerical solution, the numerical dissipation of the scheme in the turbulence simulation is quantitatively evaluated.
As a mesoscopic method, HGKS performs better than both lattice Boltzmann
method (LBM) and discrete unified gas-kinetic scheme (DUGKS), due to its high-order accuracy.
Meanwhile, based on the kinetic formulation HGKS shows advantage for supersonic turbulent flow simulation with
its accuracy and robustness.  The current work demonstrates the capability of HGKS
as a powerful DNS tool from the low speed to supersonic turbulence study, which is
less reported under the framework of finite volume scheme.
\end{abstract}
\begin{keyword}
High-order gas-kinetic scheme, direct numerical simulation of
turbulence, parallel computation.
\end{keyword}

\maketitle

\section{Introduction}
Turbulence is ubiquitous in natural phenomena and
engineering fluid applications \cite{tennekes1972first,pope2001turbulent}.
The understanding and prediction of multiscale turbulent flow is one of the most difficult problems
in both mathematics and physical sciences.
Direct numerical simulation (DNS)  solves the Navier-Stokes
equations directly, resolve all scales of the turbulent motion (above Kolmogorov scale),
and eliminate modeling entirely \cite{kim1987turbulence, moin1998direct}.
With the advances of numerical methods and super computers, great success
has been achieved by DNS to accurately compute the unsteady
turbulent flow, such as DNS of turbulent channel flow up to $Re_{\tau} \approx 5200$ \cite{lee2015direct}.
For the incompressible turbulence simulation, the
 spectral and pseudo-spectral method  \cite{moin1998direct, wang1996examination}, and lattice Boltzmann method (LBM) \cite{chen1998lattice, yu2005lattice}
have been established and validated successfully. However, for the compressible flow simulation with
with discontinuous shocks \cite{GKS-high-4}, both of them fail to
capture shocklets and suffer from numerical instability. For the
compressible turbulence flow \cite{sarkar1991analysis,sagaut2008homogeneous}, the high-order finite
difference WENO scheme \cite{WENO-liu, WENO-JS, WENO-Z} and the high-order
compact method \cite{lele1992compact} have been widely utilized.
Aiming at capturing shocklets robustly and resolving smooth region
accurately, the hybrid scheme combining the compact scheme and WENO
scheme has been developed \cite{wang2010hybrid}. However, due to the
numerical instability when encountering strong shocklets, the highest
turbulent Mach number for hybrid scheme is still limited and critical threshold of simulating supersonic flow remains.
Although the second-order finite volume scheme is the main workhorse in practical engineering applications,
the DNS is reported rarely within the finite volume framework due to its over-dissipative nature \cite{slotnick2014cfd}.
Because of the advantage of the finite volume formulation, such as the excellent conservative properties and favorable ability in  capturing discontinuities, 
it is reasonable to develop high-order finite volume scheme for direct simulation of turbulent flow in all flow regimes
from subsonic to supersonic ones.

In the past decades, the gas-kinetic scheme (GKS) has been
developed systematically based on the Bhatnagar-Gross-Krook (BGK)
model \cite{BGK-1,BGK-2} under the finite volume framework, and applied
successfully for the computations from low speed flow to hypersonic one
\cite{GKS-Xu1,GKS-Xu2}. Different from the classical methods with
Riemann solvers \cite{Riemann-appro}, the gas-kinetic scheme
presents a gas evolution process from kinetic scale to hydrodynamic
scale, where both inviscid and viscous fluxes are recovered from a
time-dependent and genuinely multi-dimensional gas distribution
function at a cell interface. In discontinuous shock region, the
kinetic scale particle transport physics takes effect to construct a crisp and stable
shock transition. In smooth flow region, the accurate Navier-Stokes
solution can be obtained once the flow structure is well resolved.
Starting from a time-dependent flux function, based on the two-stage fourth-order
formulation \cite{GRP-high-1,GRP-high-2}, a high-order gas-kinetic scheme has been constructed and applied for the
compressible flow simulation \cite{GKS-high-1,GKS-high-2,GKS-high-3}.
The fourth-order and even higher-order can be achieved in GKS with the implementation of the traditional
second-order or third-order GKS evolution model.
More importantly, the high-order GKS is as robust as the second-order scheme and works perfectly
from the subsonic to hypersonic viscous heat conducting flows \cite{GKS-high-5}.
In recent years, the gas-kinetic scheme has been applied in the
turbulent flow simulation successfully as well. For high-Reynolds number
turbulent flow, the gas-kinetic scheme coupled with traditional
eddy viscosity turbulence model has been developed and implemented
in turbulent flow study \cite{GKS-tub-1,GKS-tub-2,GKS-tub-3}.

Recently, with the implementation of two-stage temporal discretization and
WENO reconstruction, high-order gas-kinetic scheme (HGKS) in three dimensional space has been
successfully developed in the DNS for isotropic compressible turbulence \cite{GKS-high-4}, which is the
first attempt by gas-kinetic scheme to the DNS study. The isotropic compressible turbulence with high
turbulent Mach number up to supersonic regime has been studied, which
verifies the validity of HGKS for compressible turbulence study, especially in the high speed regime.
In order to resolve the small-scale flow structure and present the results at high Reynolds number,
the development of a parallel HGKS is necessary.
Here, the domain decomposition and message passing interface (MPI) \cite{MPI} will be implemented in HGKS.
Then, the classical turbulent tests from nearly incompressible flow to the hypersonic one, including Taylor-Green
vortex, turbulent channel flow, and isotropic compressible turbulence, are used to validate the parallel scalability,
efficiency, accuracy, and robustness of HGKS.
The performance of HGKS is compared with the classical methods,
including the popular high-order finite difference scheme \cite{Case-Debonis}, lattice Boltzmann
method (LBM), and discrete unified gas-kinetic scheme (DUGKS)
\cite{DUGKS}. For the nearly incompressible flows, the performance
of HGKS is comparable with the finite difference scheme,
including the resolution and efficiency of
computation. As a mesoscopic method, HGKS performs better than both LBM and DUGKS.
Meanwhile, HGKS shows advantage for supersonic
turbulence study due to its accuracy and robustness. For the isotropic compressible
 turbulence, the cases with high turbulent Mach number can
be simulated without any special treatment.
The current study provides us confidence on the further investigation of
compressible turbulence, such as shock-boundary
interaction and supersonic turbulent boundary layer transition.

This paper is organized as follows. In Section 2, the high-order
gas-kinetic scheme and the strategy of parallelization are
introduced. Section 3 includes numerical simulation and discussions.
The last section is the conclusion.

\section{High-order GKS and parallel implementation}
\subsection{High-order GKS}
The three-dimensional BGK equation \cite{BGK-1,BGK-2} can be
written as
\begin{equation}\label{bgk}
f_t+uf_x+vf_y+wf_z=\frac{g-f}{\tau},
\end{equation}
where $\boldsymbol{u}=(u,v,w)$ is the particle velocity, $f$ is the
gas distribution function, $g$ is the three-dimensional Maxwellian
distribution and $\tau$ is the collision time. The collision term
satisfies the compatibility condition
\begin{equation} \label{compatibility}
\int \frac{g-f}{\tau}\psi \text{d}\Xi=0,
\end{equation}
where
$\displaystyle\psi=(1,u,v,w,\frac{1}{2}(u^2+v^2+w^2+\xi^2))^T$, $\xi^2=\xi_1^2+...+\xi_N^2$,
$\text{d}\Xi=\text{d}u\text{d}v\text{d}w\text{d}\xi_1…\text{d}\xi_{N}$,
$\gamma$ is the specific heat ratio and  $N=(5-3\gamma)/(\gamma-1)$
is the internal degree of freedom.

Taking moments of the BGK equation Eq.\eqref{bgk} and integrating
with respect to space, the finite volume scheme can be expressed as
\begin{align}\label{semi}
\frac{\text{d}(Q_{ijk})}{\text{d}t}=\mathcal{L}(Q_{ijk}),
\end{align}
where the operator $\mathcal{L}$ is defined as
\begin{equation}\label{finite}
\mathcal{L}(Q_{ijk})=-\frac{1}{|\Omega_{ijk}|}\sum_{p=1}^6\mathbb{F}_{p}(t),
\end{equation}
where $\Omega_{ijk}=\overline{x}_i\times\overline{y}_j\times
\overline{z}_k$ with $\overline{x}_i=[x_i-\Delta x/2,x_i+\Delta
x/2], \overline{y}_j=[y_j-\Delta y/2,y_j+\Delta y/2],
\overline{z}_k=[z_k-\Delta z/2,z_k+\Delta z/2]$, $\mathbb{F}_{p}(t)$
is the numerical flux across the cell interface $\Sigma_{p}$. The
numerical flux in $x$-direction is given as example
\begin{align*}
\mathbb{F}_{p}(t)=\iint_{\Sigma_{p}}
F(Q)\cdot\boldsymbol{n}\text{d}\sigma=\sum_{m,n=1}^2\omega_{mn}
\int\psi u
f(\boldsymbol{x}_{i+1/2,j_m,k_n},t,\boldsymbol{u},\xi)\text{d}\Xi\Delta y\Delta z,
\end{align*}
where $\boldsymbol{n}$ is the outer normal direction.
In this paper, the orthogonal Cartesian mesh is considered, the normal direction is constant for each cell interface.
The Gaussian
quadrature is used over the cell interface, where $\omega_{mn}$ is
the quadrature weight,
$\boldsymbol{x}_{i+1/2,m,n}=(x_{i+1/2},y_{j_m},z_{k_n})$ and
$(y_{j_m},z_{k_n})$ is the Gauss quadrature point of cell
interface $\overline{y}_j\times\overline{z}_k$. The gas distribution
function $f(\boldsymbol{x}_{i+1/2,j_m,k_n},t,\boldsymbol{u},\xi)$ in the local coordinate
can be given by the integral solution of BGK equation Eq.\eqref{bgk}
as follows
\begin{equation*}
f(\boldsymbol{x}_{i+1/2,j_m,k_n},t,\boldsymbol{u},\xi)=\frac{1}{\tau}\int_0^t
g(\boldsymbol{x}',t',\boldsymbol{u},\varsigma)e^{-(t-t')/\tau}\text{d}t'+e^{-t/\tau}f_0(-\boldsymbol{u}t,\xi),
\end{equation*}
where $\boldsymbol{u}=(u,v,w)$ is the particle velocity,
$\boldsymbol{x}'=\boldsymbol{x}_{i+1/2,j_m,k_n}-\boldsymbol{u}(t-t')$
is the trajectory of particles, $f_0$ is the initial gas
distribution function, and $g$ is the corresponding equilibrium
state. With the first order spatial derivatives, the second-order
gas distribution function at cell interface can be expressed as
\begin{align}\label{flux}
f(\boldsymbol{x}_{i+1/2,j_m,k_n},t,\boldsymbol{u},\xi)=&(1-e^{-t/\tau})g_0+
((t+\tau)e^{-t/\tau}-\tau)(\overline{a}_1u+\overline{a}_2v+\overline{a}_3w)g_0\nonumber\\
+&(t-\tau+\tau e^{-t/\tau}){\bar{A}} g_0\nonumber\\
+&e^{-t/\tau}g_r[1-(\tau+t)(a_{1}^{r}u+a_{2}^{r}v+a_{3}^{r}w)-\tau A^r)]H(u)\nonumber\\
+&e^{-t/\tau}g_l[1-(\tau+t)(a_{1}^{l}u+a_{2}^{l}v+a_{3}^{l}w)-\tau
A^l)](1-H(u)),
\end{align}
where the equilibrium state $g_{0}$ and the corresponding conservative
variables $Q_{0}$ can be determined by the compatibility
condition
\begin{align*}
\int\psi g_{0}\text{d}\Xi=Q_0=\int_{u>0}\psi
g_{l}\text{d}\Xi+\int_{u<0}\psi g_{r}\text{d}\Xi.
\end{align*}
With the reconstruction of macroscopic variables, the coefficients in Eq.\eqref{flux} can be fully determined by
the reconstructed derivatives and compatibility
condition
\begin{align*}
\displaystyle \langle a_{1}^{k}\rangle=\frac{\partial
Q_{k}}{\partial x}, \langle
a_{2}^{k}\rangle=\frac{\partial Q_{k}}{\partial y},
\langle a_{3}^{k}\rangle&=\frac{\partial Q_{k}}{\partial
z}, \langle
a_{1}^{k}u+a_{2}^{k}v+a_{3}^{k}w+A^{k}\rangle=0,\\ \displaystyle
\langle\overline{a}_1\rangle=\frac{\partial Q_{0}}{\partial
x}, \langle\overline{a}_2\rangle=\frac{\partial
Q_{0}}{\partial y},
\langle\overline{a}_3\rangle&=\frac{\partial Q_{0}}{\partial
z},
\langle\overline{a}_1u+\overline{a}_2v+\overline{a}_3w+\overline{A}\rangle=0,
\end{align*}
where $k=l,r$ and $\langle...\rangle$ are the moments of the
equilibrium $g$ and defined by
\begin{align*}
\langle...\rangle=\int g (...)\psi \text{d}\Xi.
\end{align*}
More details of the gas-kinetic scheme can be found in the literatures
\cite{GKS-Xu1,GKS-Xu2,GKS-Xu3}. Thus, the gas distribution function is determined, and the numerical flux can be obtained by taking moments of it.
For the high-order spatial accuracy, the fifth-order WENO method
\cite{WENO-JS, WENO-Z} is adopted. For the
three-dimensional computation, the dimension-by-dimension
reconstruction is used. More details about spatial reconstruction
can be found in previous work \cite{GKS-high-1,GKS-high-4}, and several remarks
are given.

\begin{rmk}
For the low speed flows, such as Taylor-Green vortex problem and turbulent channel
flow, the flow fields are smooth without strong shocklets, the
simplified smooth second-order gas-kinetic flux \cite{GKS-high-3}
and WENO scheme with linear weights are used to avoid the
numerical dissipation from artificially created interface discontinuity. For the tangential reconstruction  $Q_{l,r}$
and $Q_0$, the fourth-order polynomials are constructed at the
horizontal and vertical direction. The variables and spatial
derivatives can be constructed at the Gaussian quadrature points.
For the compressible isotropic turbulence from subsonic to
supersonic regime, the WENO-Z \cite{WENO-Z} scheme is used. In order
to eliminate the spurious oscillation and improve the stability, the
reconstruction can be performed for the characteristic variables in
local coordinate for each Gaussian quadrature point.
The characteristic variable is defined as $\omega=R^{-1}Q$, where $Q$ is variable in
the local coordinate, and where $R$ is the right eigenmatrix of Jacobian matrix
$(\partial F/\partial Q)_{G}$ at Gaussian quadrature point. With the
reconstructed variable, the conservative variables can be obtained
by the inverse projection.
For the tangential reconstruction of
$Q_{l,r}$ and $Q_0$, the variables at the ends of cell interface can
be obtained from the fifth-order WENO method at the horizontal and
vertical direction. With reconstructed variables and the cell
averaged variables, the quadratic polynomials can be constructed.
The variables and spatial derivatives can be constructed at the
Gaussian quadrature points as well.
\end{rmk}

Based on the time-dependent flux function of the generalized Riemann
problem solver (GRP) \cite{GRP-high-1,GRP-high-2} and gas-kinetic scheme, a two-stage
fourth-order time-accurate discretization was developed for
Lax-Wendroff type flow solvers
\cite{GKS-high-1,GKS-high-2}.
Consider the following time dependent equation
\begin{align*}
\frac{\partial Q}{\partial t}=\mathcal {L}(Q),
\end{align*}
with the initial condition at $t_n$, i.e.,
\begin{align*}
Q(t=t_n)=Q^n,
\end{align*}
where $\mathcal {L}$ is an operator for spatial derivative of flux,
the state $Q^{n+1}$ at $t_{n+1}=t_n+\Delta t$  can be updated with
the following formula
\begin{equation}\label{two-stage}
\begin{split}
&Q^*=Q^n+\frac{1}{2}\Delta t\mathcal {L}(Q^n)+\frac{1}{8}\Delta
t^2\partial_t\mathcal{L}(Q^n), \\
Q^{n+1}=&Q^n+\Delta t\mathcal {L}(Q^n)+\frac{1}{6}\Delta
t^2\big(\partial_t\mathcal{L}(Q^n)+2\partial_t\mathcal{L}(Q^*)\big).
\end{split}
\end{equation}
It can be proved that for hyperbolic equations the above temporal
discretization provides a fourth-order time accurate solution for
$Q^{n+1}$. To implement two-stage fourth-order method for
Eq.\eqref{semi}, a linear function is used to approximate the time
dependent numerical flux
\begin{align}\label{expansion-1}
\mathbb{F}_{p}(t)\approx\mathbb{F}_{p}^n+ \partial_t
\mathbb{F}_{p}^n(t-t_n).
\end{align}
Integrating Eq.\eqref{expansion-1} over $[t_n, t_n+\Delta t/2]$ and
$[t_n, t_n+\Delta t]$, we have the following two equations
\begin{align*}
\mathbb{F}_{p}^n\Delta t&+\frac{1}{2}\partial_t
\mathbb{F}_{p}^n\Delta t^2 =\int_{t_n}^{t_n+\Delta t}\mathbb{F}_{p}(t)\text{d}t, \\
\frac{1}{2}\mathbb{F}_{p}^n\Delta t&+\frac{1}{8}\partial_t
\mathbb{F}_{p}^n\Delta t^2 =\int_{t_n}^{t_n+\Delta t/2}\mathbb{F}_{p}(t)\text{d}t.
\end{align*}
The coefficients $\mathbb{F}_{p}^n$ and $\partial_t
\mathbb{F}_{p}^n$  at the initial stage can be determined by solving
the linear system. According to Eq.\eqref{finite}, $\mathcal
{L}(Q_{i}^n)$ and the temporal derivative $\partial_t\mathcal
{L}(Q_{i}^n)$ at $t^n$ can be constructed by
\begin{align*}
\mathcal{L}(Q_{i}^n)&=-\frac{1}{|\Omega_{i}|}\sum_{p=1}^6\mathbb{F}_{p}^n,~~
\partial_t\mathcal{L}(Q_{i}^n)=-\frac{1}{|\Omega_{i}|}\sum_{p=1}^6\partial_t\mathbb{F}_{p}^n.
\end{align*}
The flow variables $Q^*$ at the
intermediate stage can be updated. Similarly,  $\mathcal
{L}(Q_{i}^*),
\partial_t\mathcal {L}(Q_{i}^*)$ at the
intermediate state can be constructed and $Q^{n+1}$ can be updated as well.

\begin{figure}[!h]
	\centering
	\includegraphics[width=0.65\textwidth]{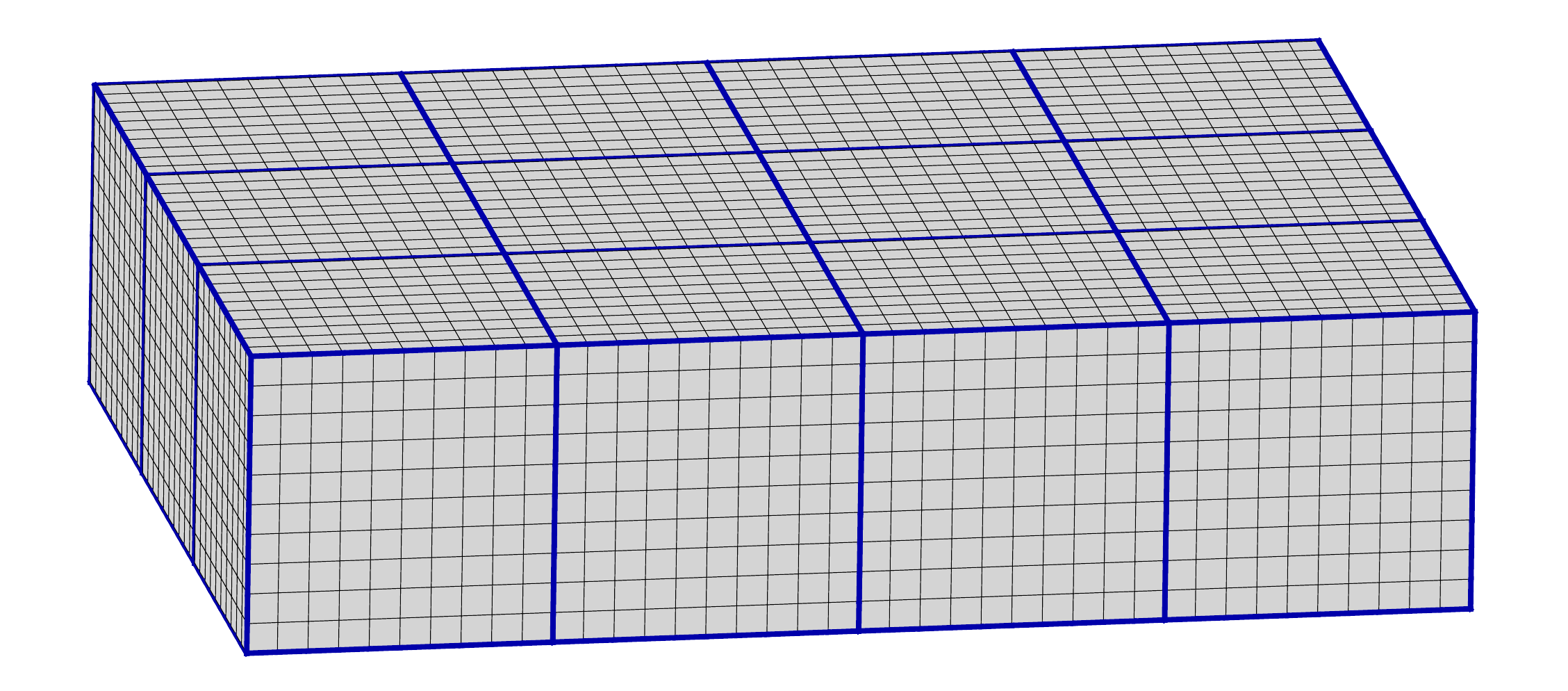}
	\caption{\label{mesh} Schematic for two-dimensional domain decomposition with $n_y=4, n_z=3$.}
\end{figure}

\subsection{Parallel implementation}
Due to the explicit formulation of HGKS, a
popular parallel strategy is developed, where two-dimensional domain
decomposition is used. As shown in Fig.\ref{mesh}, the total number of cells is $N_x\times N_y\times N_z$, and the computational
domain is divided into $n_y$ parts in $y$-direction, $n_z$ parts in
$z$-direction and no division is used in $x$-direction. The
processor $P_{jk}, j=0,…,n_y-1, k=0,...,n_z-1$ handles a
sub-domain with $N_x \times ny_j\times nz_k$ cells, where
\begin{align*}
\begin{cases}
ny_j=[N_y/n_y]+1, &j<\text{mod}(N_y,n_y),\\
ny_j=[N_y/n_y], &j\geq \text{mod}(N_y,n_y),
\end{cases}
\end{align*}
and
\begin{align*}
\begin{cases}
nz_k=[N_z/n_z]+1, &k<\text{mod}(N_z,n_z),\\
nz_k=[N_z/n_z], &k\geq \text{mod}(N_z,n_z).
\end{cases}
\end{align*}
The data communication is performed between eight neighboring sub-domains, and the specific boundary conditions are performed for the boundary processor. The procedure is
the only data communication of the algorithm, which is handled by
the MPI libraries \cite{MPI}.

\section{Numerical simulation and discussion}
In this section, numerical tests from the nearly incompressible flow to the
supersonic one will be
presented to validate our numerical scheme. For the smooth flow
without discontinuities, the collision time takes
\begin{align*}
\tau=\frac{\mu}{p}.
\end{align*}
For the flow with discontinuities, we have
\begin{align*}
\tau=\frac{\mu}{p}+C \displaystyle|\frac{p_l-p_r}{p_l+p_r}|\Delta t,
\end{align*}
where $p_l$ and $p_r$ denote the pressure on the left and right
sides of the cell interface, $\mu$ is the dynamic viscous
coefficient, $C = 1$ and $p$ is the pressure at the cell interface. The
reason for including artificial dissipation through the additional
term in the particle collision time is to enlarge the kinetic scale
physics in the discontinuous region for the construction of a
numerical shock structure through the particle free transport and
inadequate particle collision in order to keep the non-equilibrium
property.

\subsection{Taylor-Green vortex}
Taylor-Green vortex is a classical problem in fluid dynamics developed to
study vortex dynamics, turbulent transition, turbulent decay and
energy dissipation process \cite{Case-Brachet, gallis2017molecular}.
It is given by a simple construction, and
contains several key physical processes including vortex stretching,
interaction and dilatation effects.
Therefore, this case becomes an
excellent case for the evaluation of turbulent flow simulation
methodologies, and has been used by many authors for high-order
method validation
\cite{Case-Debonis,Case-Bull}. The flow is
computed within a periodic square box defined as $-\pi L\leq x, y,
z\leq \pi L$. With a uniform temperature, the initial condition is
given by
\begin{align*}
U=&V_0\sin(\frac{x}{L})\cos(\frac{y}{L})\cos(\frac{z}{L}),\\
V=&-V_0\cos(\frac{x}{L})\sin(\frac{y}{L})\cos(\frac{z}{L}),\\
W=&0,\\
p=&p_0+\frac{\rho_0V_0^2}{16}(\cos(\frac{2x}{L})+\cos(\frac{2y}{L}))(\cos(\frac{2z}{L})+2).
\end{align*}
In the computation, $L=1, V_0=1, \rho_0=1$, and the Mach number
takes $M_0=V_0/c_0=0.1$, where $c_0$ is the sound speed.  The fluid
is a perfect gas with $\gamma=1.4$, Prandtl number is $Pr=0.71$, and
Reynolds number $Re=1600$.  The characteristic convective time $t_c
= L/V_0$.
This problem is aimed at the performance of high-order gas-kinetic scheme on the
direct numerical simulation of nearly incompressible turbulent
flows. In the computation, the cases $TG_1$, $TG_2$, $TG_3$ and $TG_4$ with
$128^3$, $256^3$, $512^3$ and $1024^3$ uniform cells are tested, and the numerical results of
BB13 dispersion relation preserving (DRP) scheme \cite{bb13} with $512^3$
cells are given as reference \cite{Case-Debonis}.
The BB13 scheme was originally developed for noise computations, in which
a high-order finite difference method equipped with fourth-stage third-order algorithm for
time discretization and 13-point stencils for spatial
discretization. The compressible Navier-Stokes equations are solved
by both high-order gas-kinetic scheme and  finite
difference method.

\begin{figure}[!h]
\centering
\includegraphics[width=0.475\textwidth]{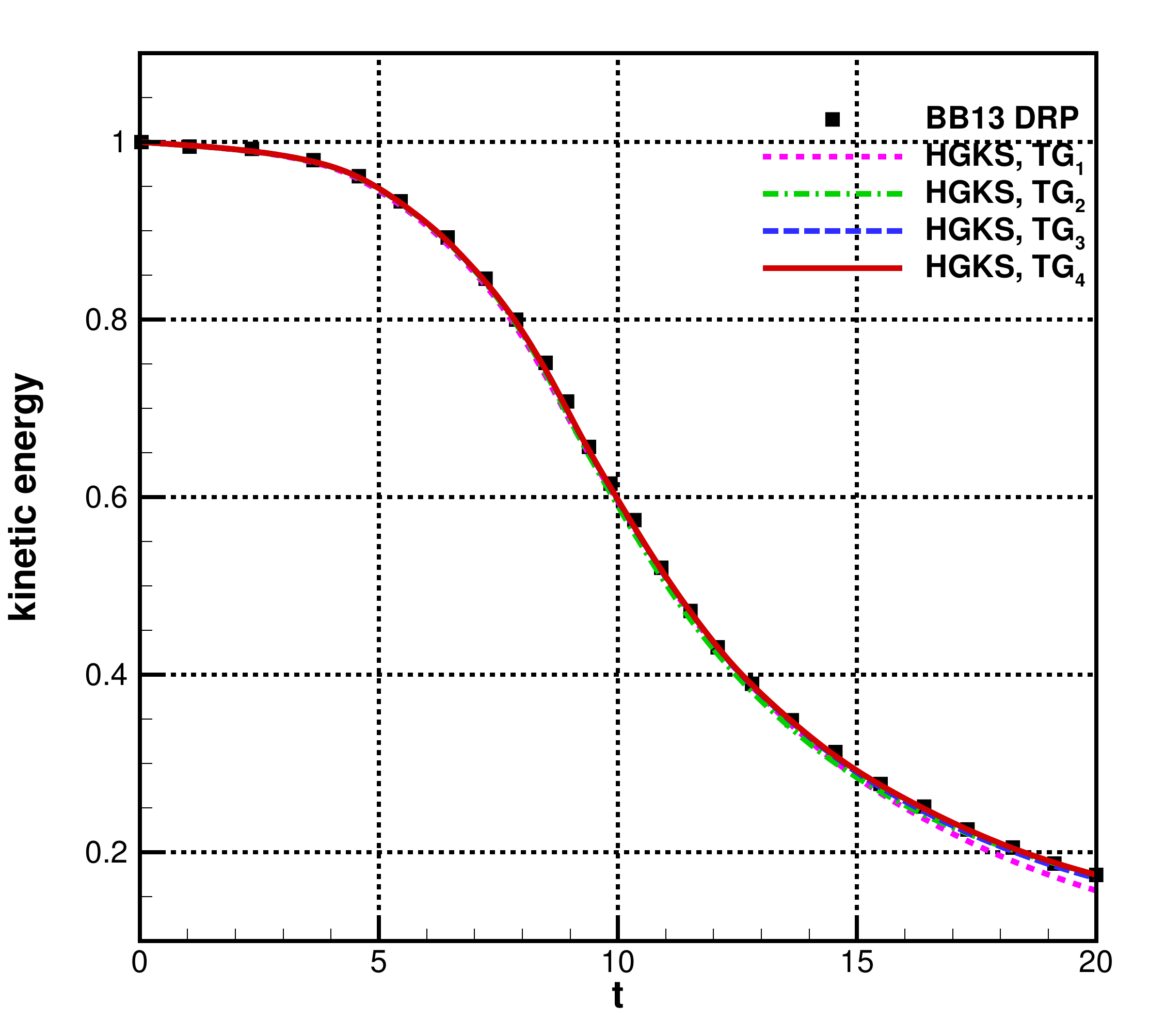}
\caption{\label{tg-vortex-1} Taylor-Green vortex: time
history of kinetic energy $E_k$.}
~\\
\includegraphics[width=0.475\textwidth]{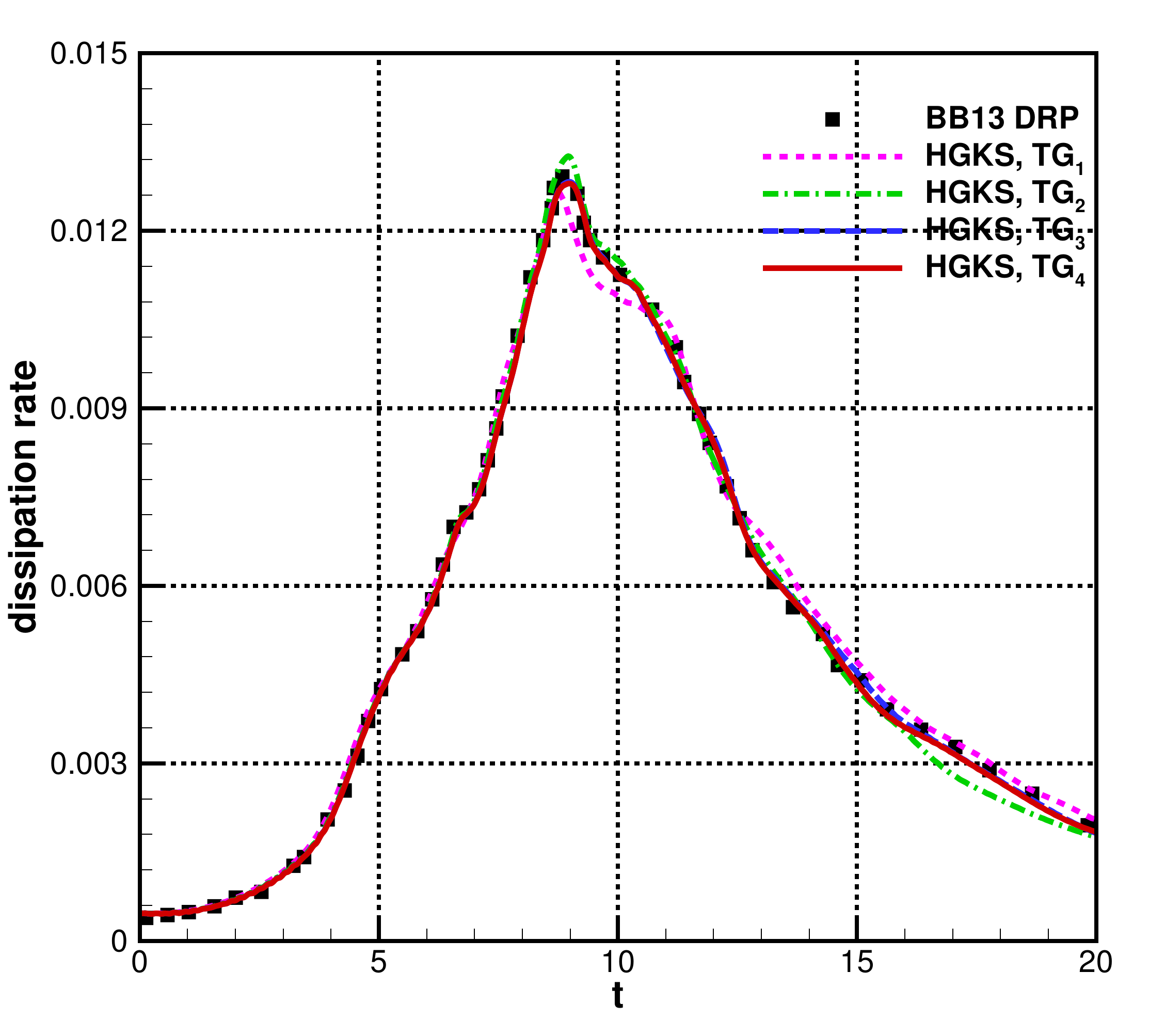}
\includegraphics[width=0.475\textwidth]{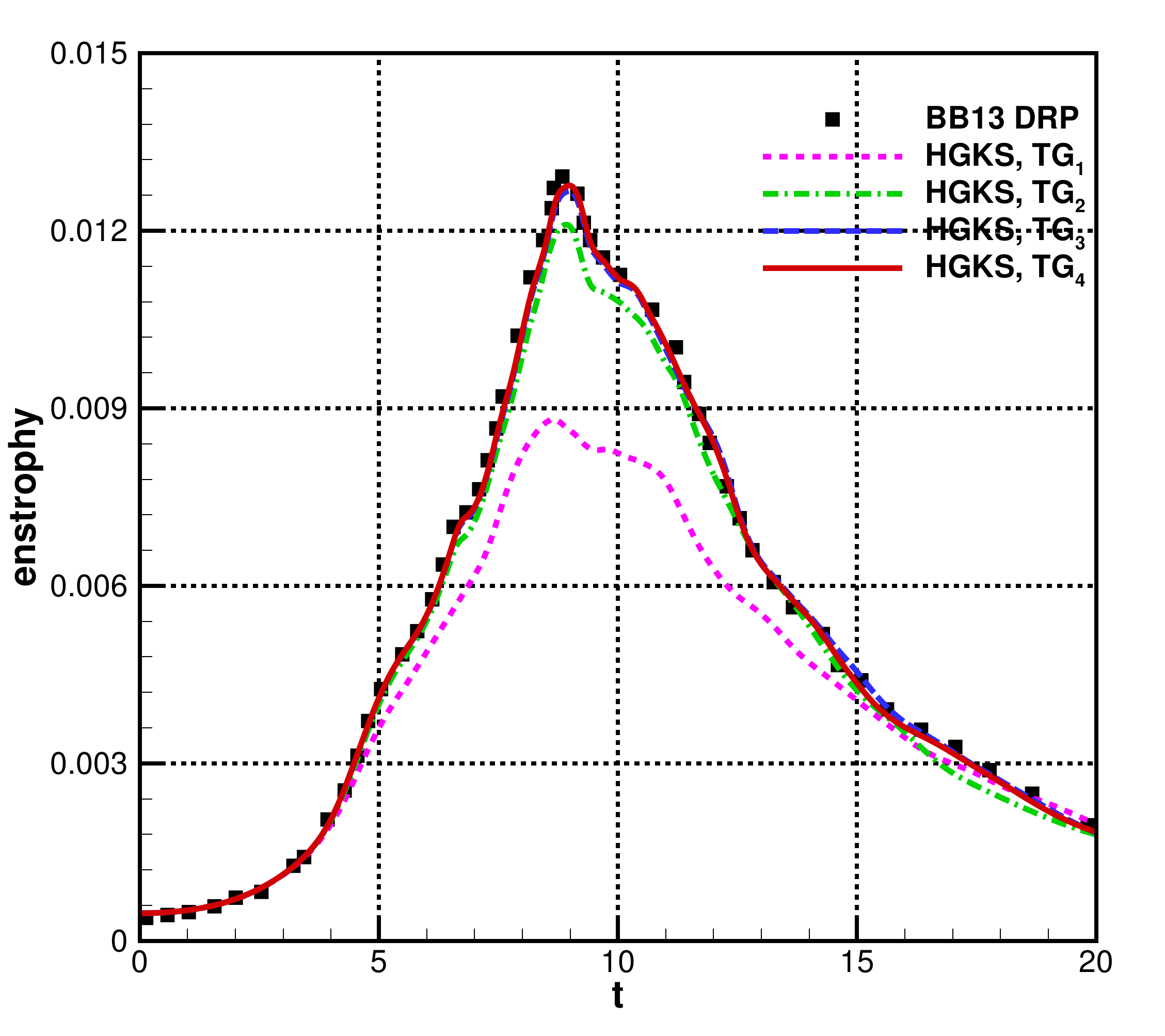}
\caption{\label{tg-vortex-2} Taylor-Green vortex: time
history of dissipation rate $\varepsilon(E_k)$ and enstrophy
$\varepsilon(\zeta)$. The reference data is $\varepsilon(E_k)$ with $512^3$ cells  for
two contours \cite{Case-Debonis}.}
\end{figure}
\begin{figure}[!h]
\centering
\includegraphics[width=0.475\textwidth]{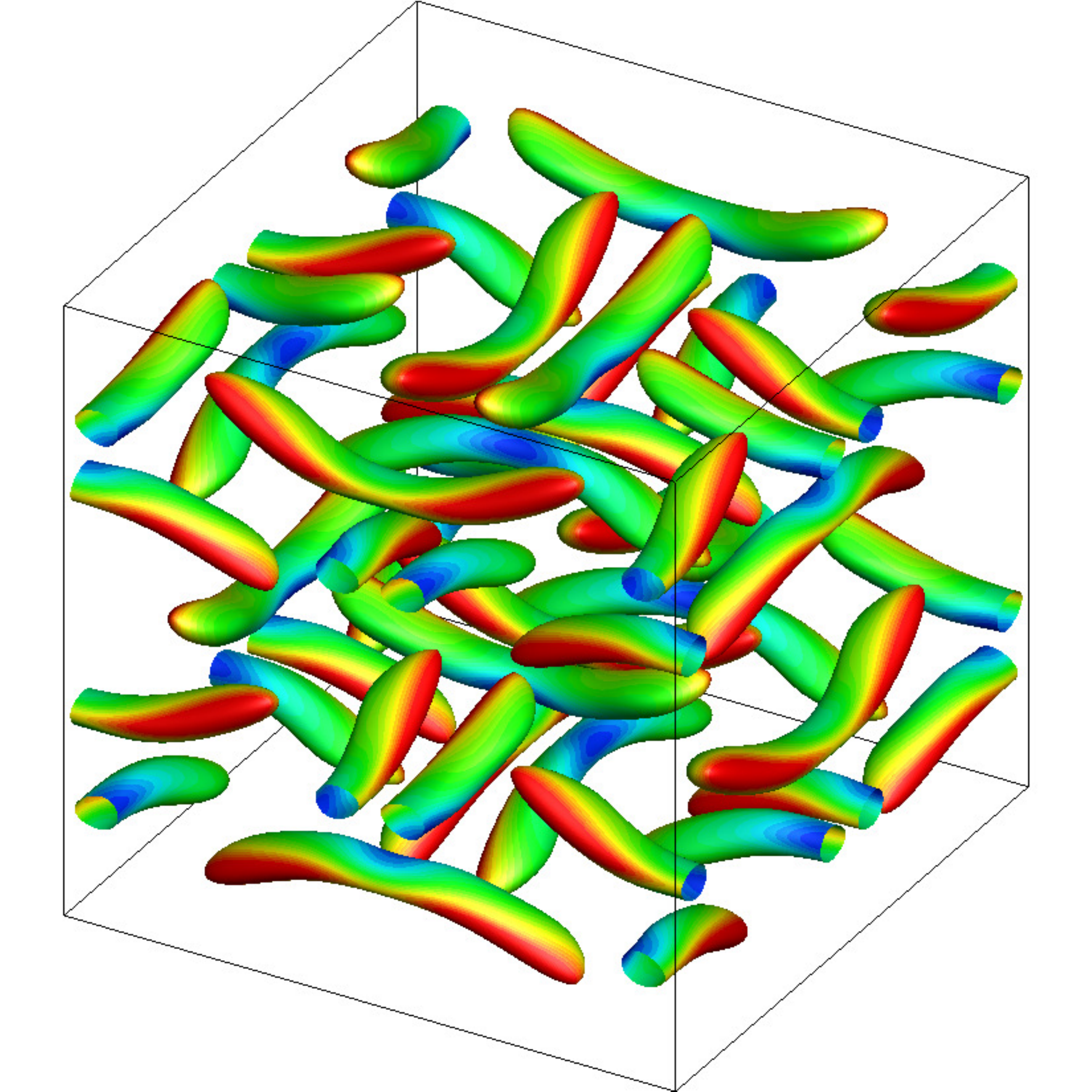}
\includegraphics[width=0.475\textwidth]{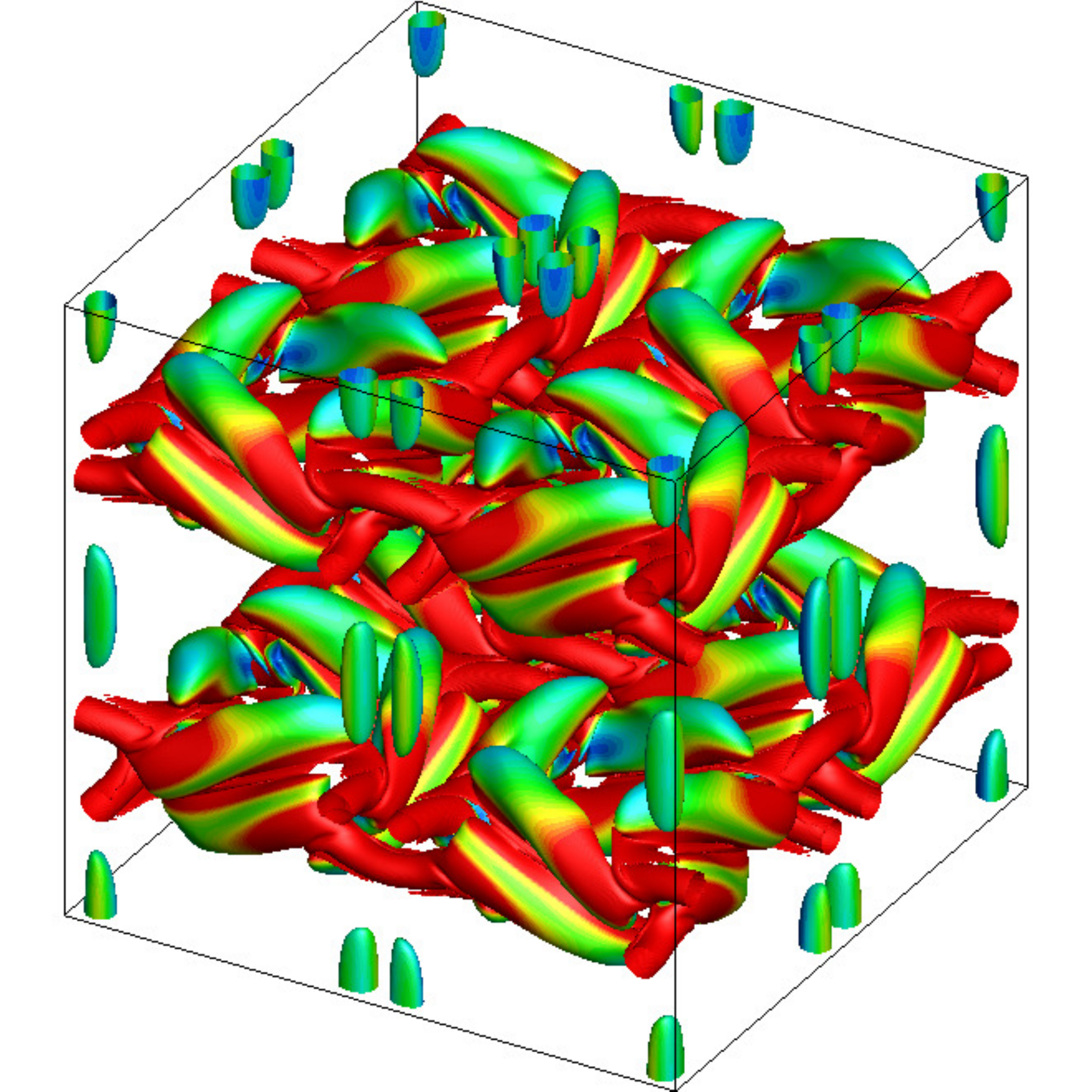}
~\\
\includegraphics[width=0.475\textwidth]{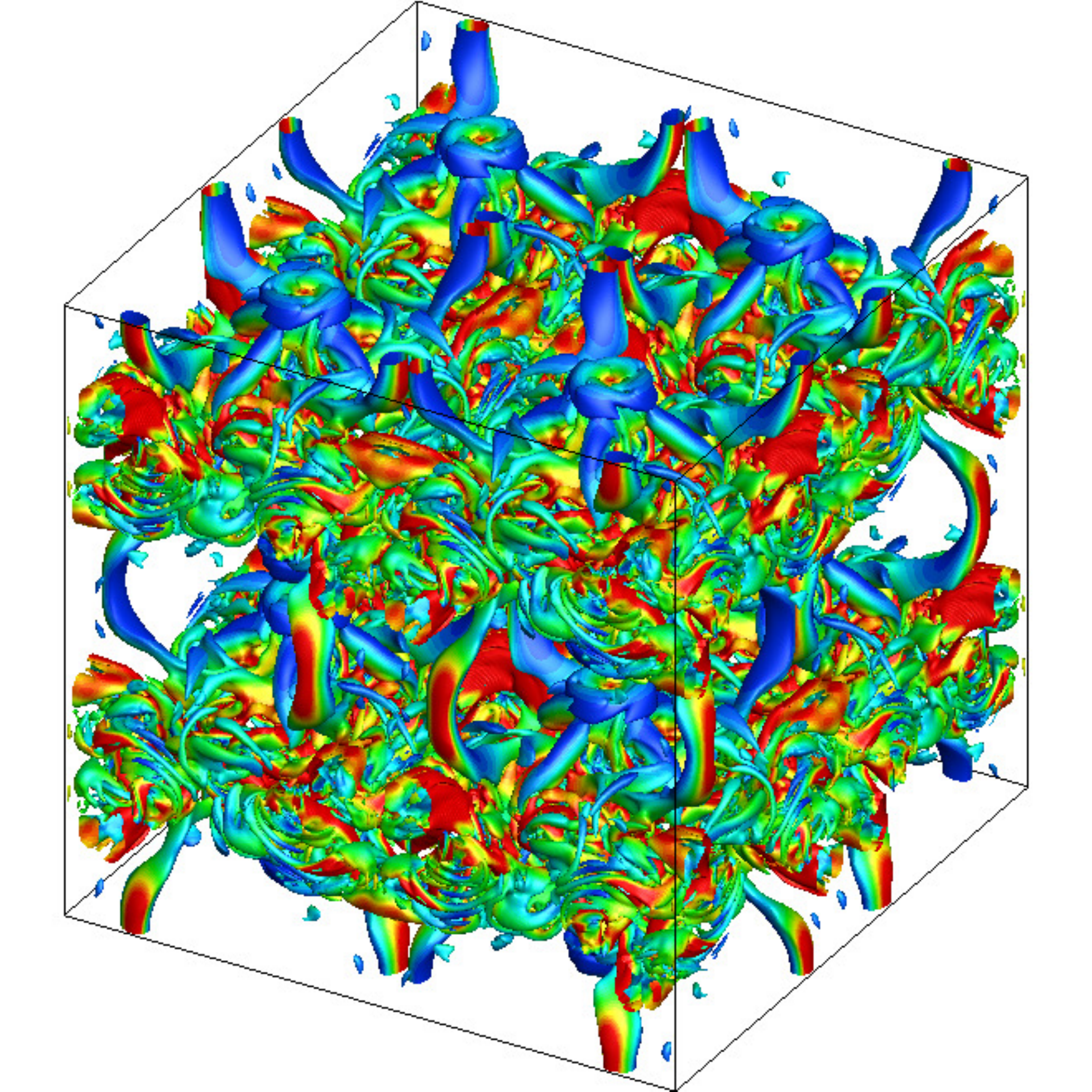}
\includegraphics[width=0.475\textwidth]{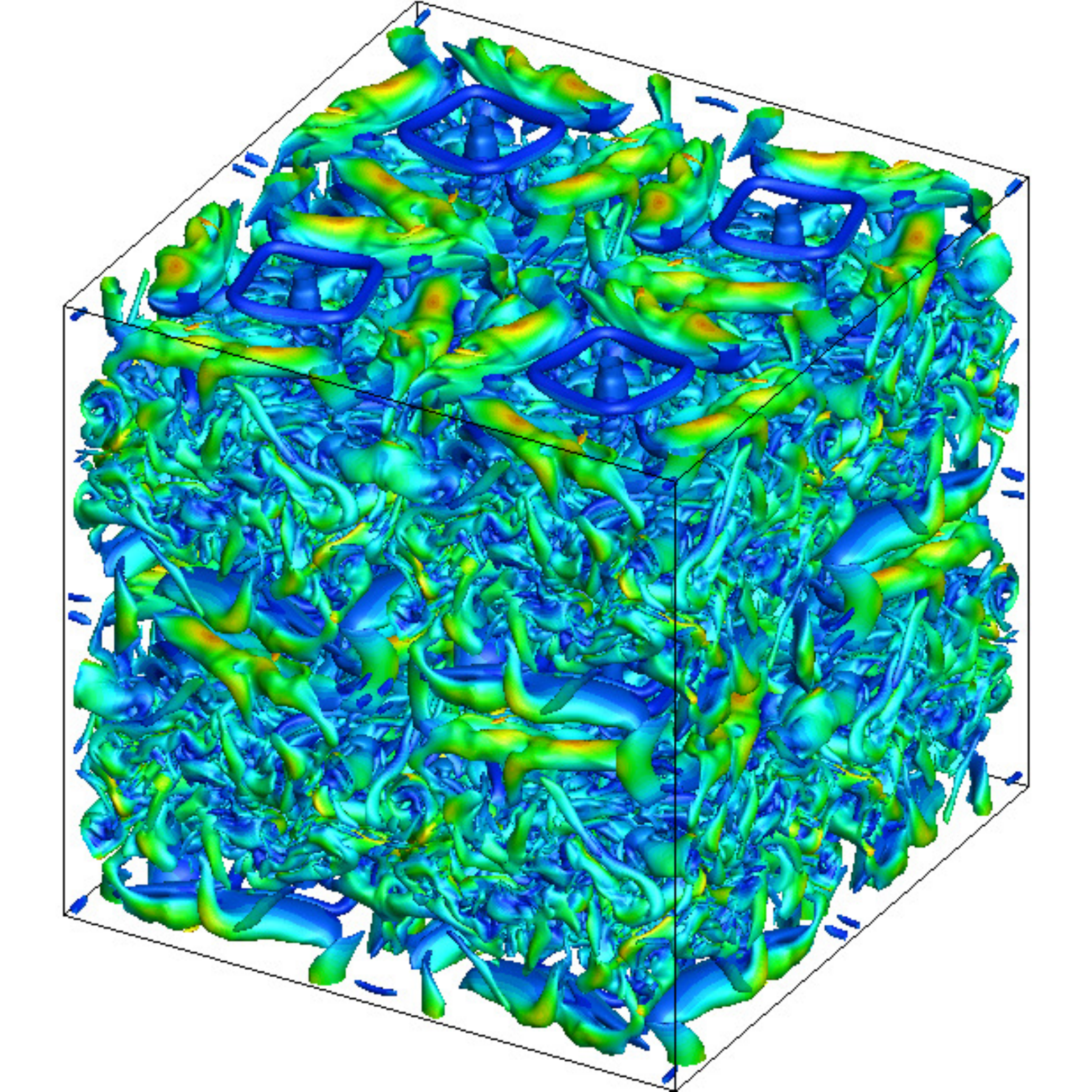}
\caption{\label{tg-vortex-4} Taylor-Green vortex:
iso-surface of the second invariant of velocity gradient tensor $Q_v =
-0.5$ at $t=2.5, 5, 10$ and $15$ colored by velocity magnitude.}
\end{figure}

To test the performance of HGKS, several diagnostic quantities
are computed from the flow as it evolves in time.
The volume-averaged kinetic energy is given by
\begin{align*}
E_k=\frac{1}{\rho_0\Omega}\int_\Omega\frac{1}{2}\rho\boldsymbol{U}\cdot\boldsymbol{U} \text{d} \Omega,
\end{align*}
where $\Omega$ is the volume of the computational domain. The
dissipation rate of kinetic energy  can be computed by the temporal
derivative of $E_k$
\begin{align*}
\varepsilon(E_k)=-\frac{\text{d}E_k}{\text{d}t},
\end{align*}
which is computed by second order central difference in the numerical results of $E_k$.
For the incompressible limit, the dissipation rate is related to the
integrated enstrophy by
\begin{align*}
\varepsilon(\zeta)=2\frac{\mu}{\rho_0}\zeta,
\end{align*}
where $\mu$ is the coefficient of viscosity, $\boldsymbol{\omega}=\nabla\times \boldsymbol{U}$ and
\begin{align*}
\zeta=\frac{1}{\rho_0\Omega}\int_\Omega\frac{1}{2}\rho\boldsymbol{\omega}\cdot\boldsymbol{\omega} \text{d} \Omega.
\end{align*}
The time history of kinetic energy is shown in
Fig.\ref{tg-vortex-1}, where the reasonable agreement is observed
with the reference solution except for the simulation with $128^3$ cells.
The kinetic energy dissipation rates $\varepsilon(E_k)$ and the enstrophy
integral computed $\varepsilon(\zeta)$ are shown in
Fig.\ref{tg-vortex-2}, respectively.
A large discrepancy is observed
in the peak dissipation rate for $\varepsilon(E_k)$ and
$\varepsilon(\zeta)$ with $128^3$ cells and $256^3$ cells, and an excellent agreement
with the reference solution is obtained with the mesh refinement.
Especially, the mesh with $1024^3$ cells is the finest
resolution for the Taylor-Green vortex problem, and a benchmark
results have been provided.
As time evolves, the vortex roll-up, stretch and
interact, eventually breaking down into turbulence.
The iso-surface of the second invariant of velocity gradient tensor $Q_v$
colored by velocity magnitude at $t = 2.5, 5, 10$ and $15$ with
$512^3$ cells are shown in Fig.\ref{tg-vortex-4}.
Velocity magnitude ranges from $0$ to $0.2$ and $20$ equivalent levels are used.
At the earliest time, the flow behaves inviscidly as the vortex begin to
evolve and roll-up. At $t=10$, the coherent structures breakdown.
Beyond this breakdown, the flow is fully turbulent and the
structures slowly decay until the flow comes to rest. The results
indicate that the resolution of HGKS is comparable even with the
higher-order finite difference method, which is widely used in DNS
of turbulence.

\begin{figure}[!h]
\centering
\includegraphics[width=0.475\textwidth]{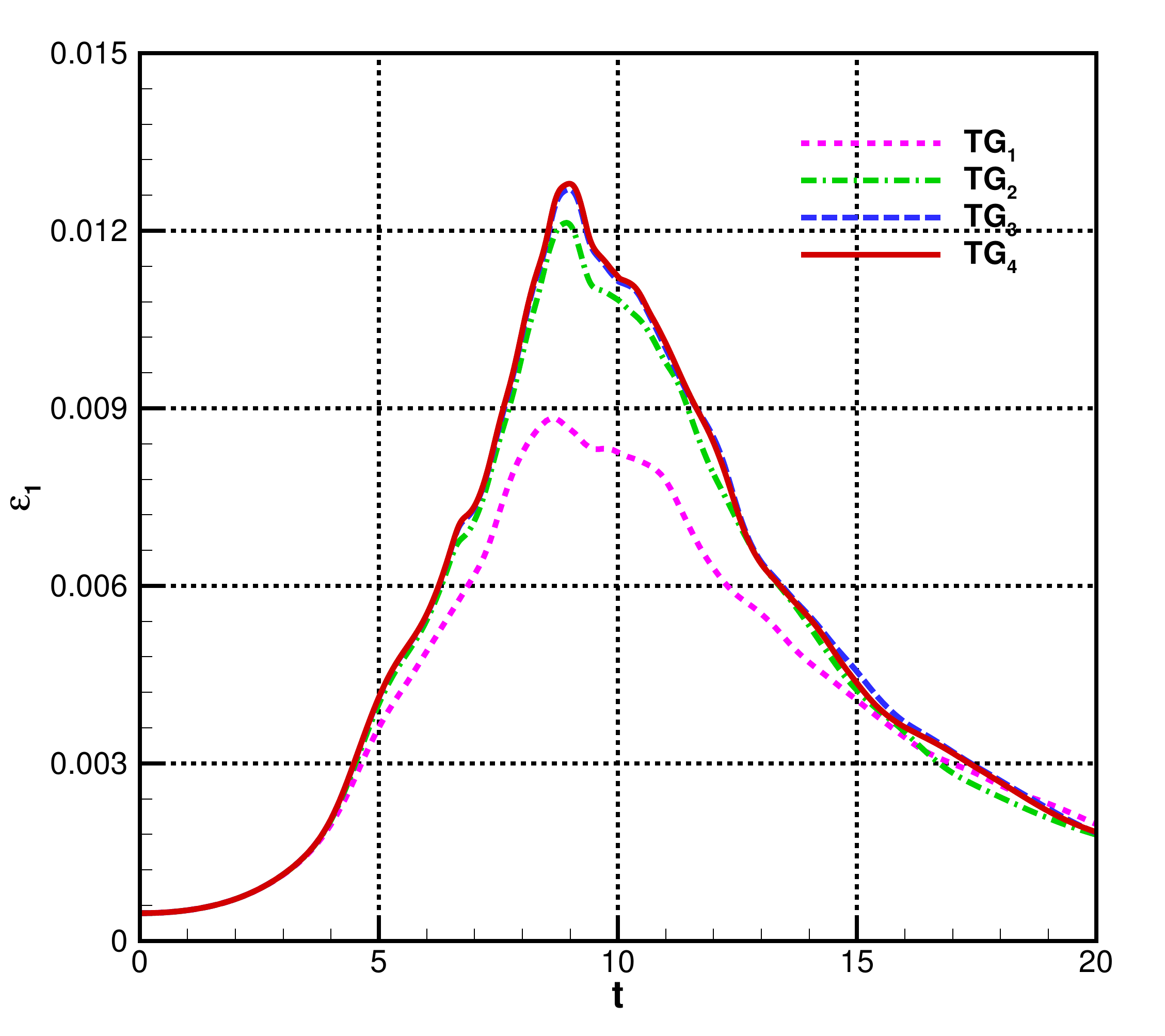}
\includegraphics[width=0.475\textwidth]{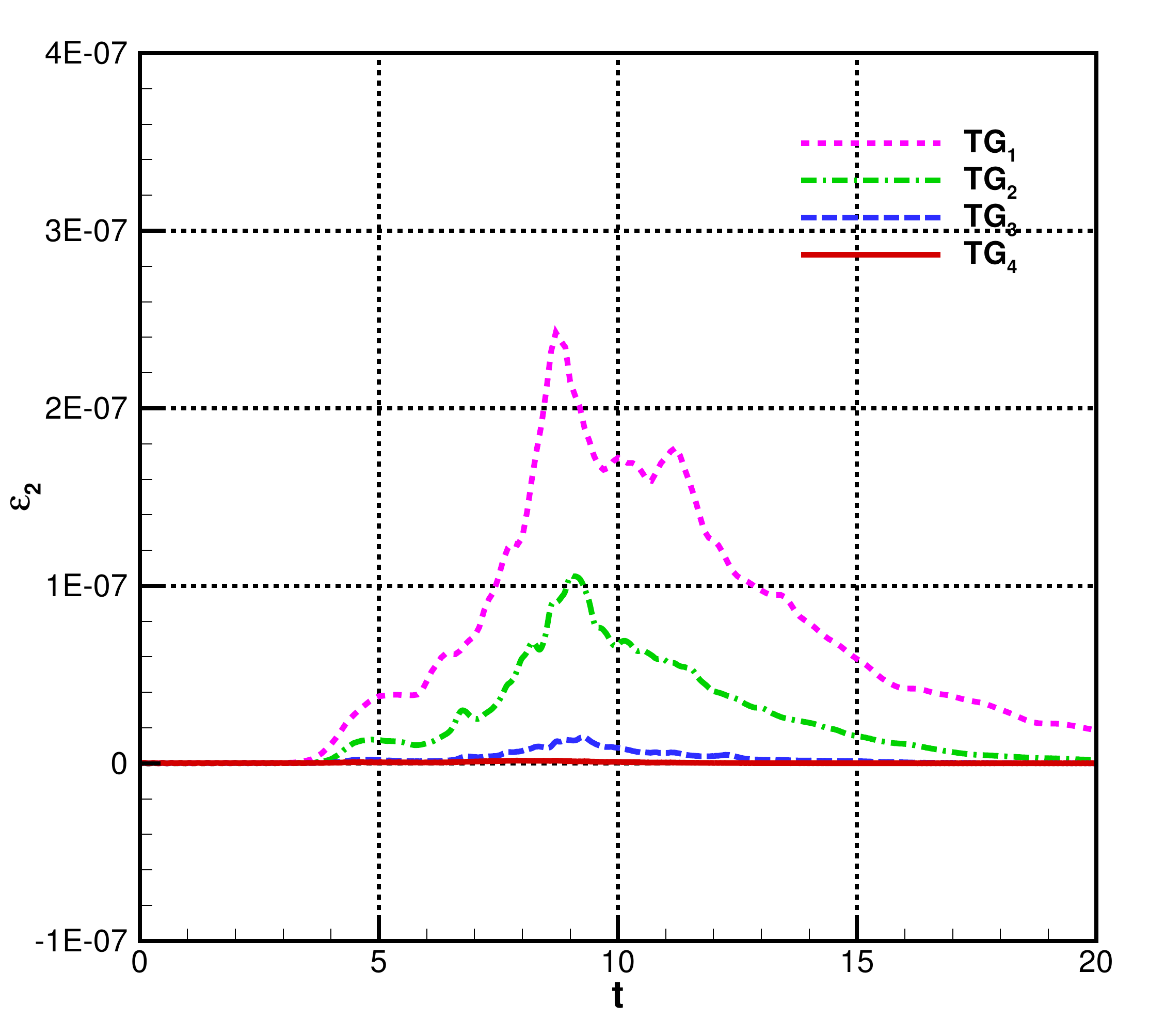}
~\\
\includegraphics[width=0.475\textwidth]{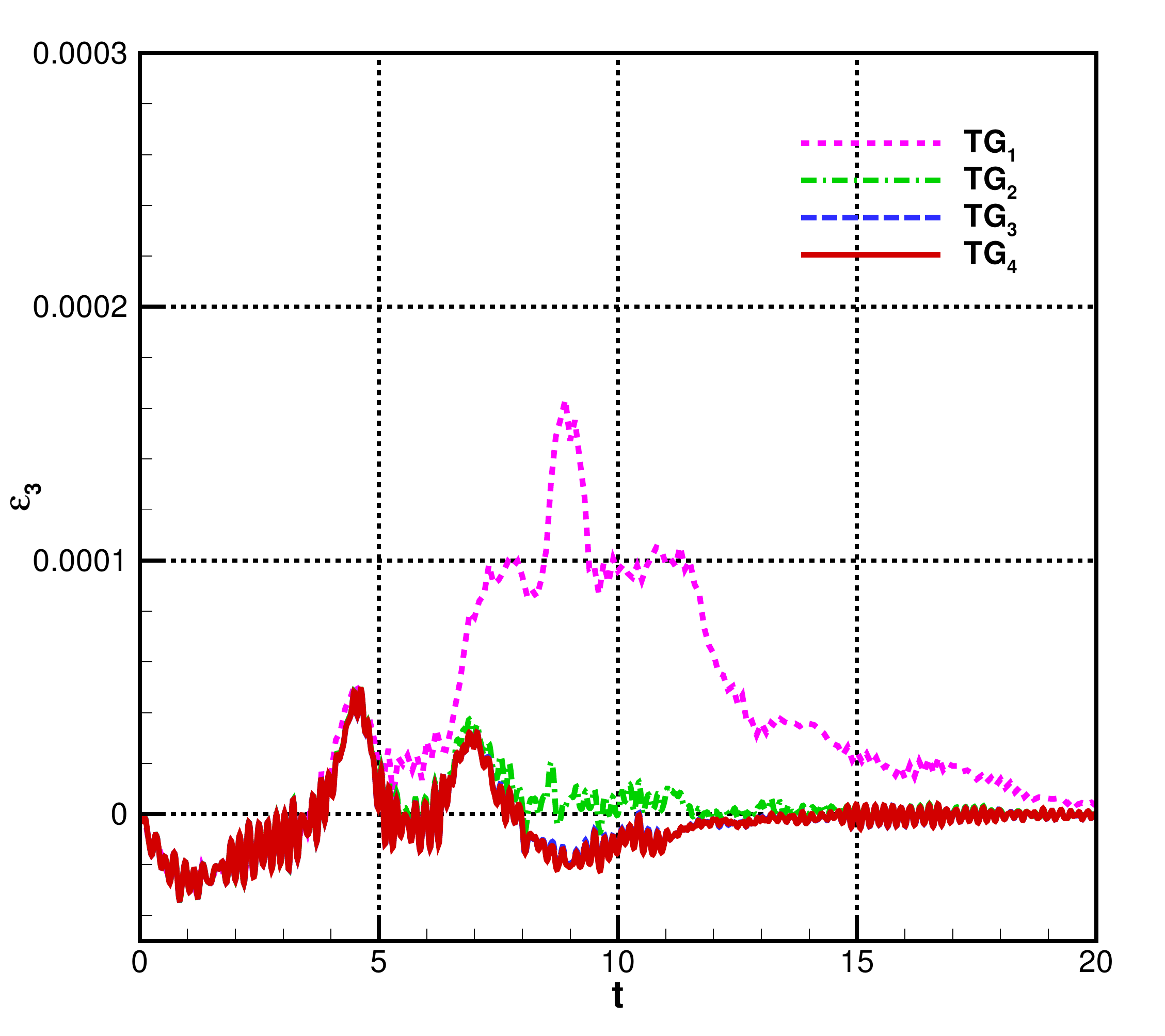}
\includegraphics[width=0.475\textwidth]{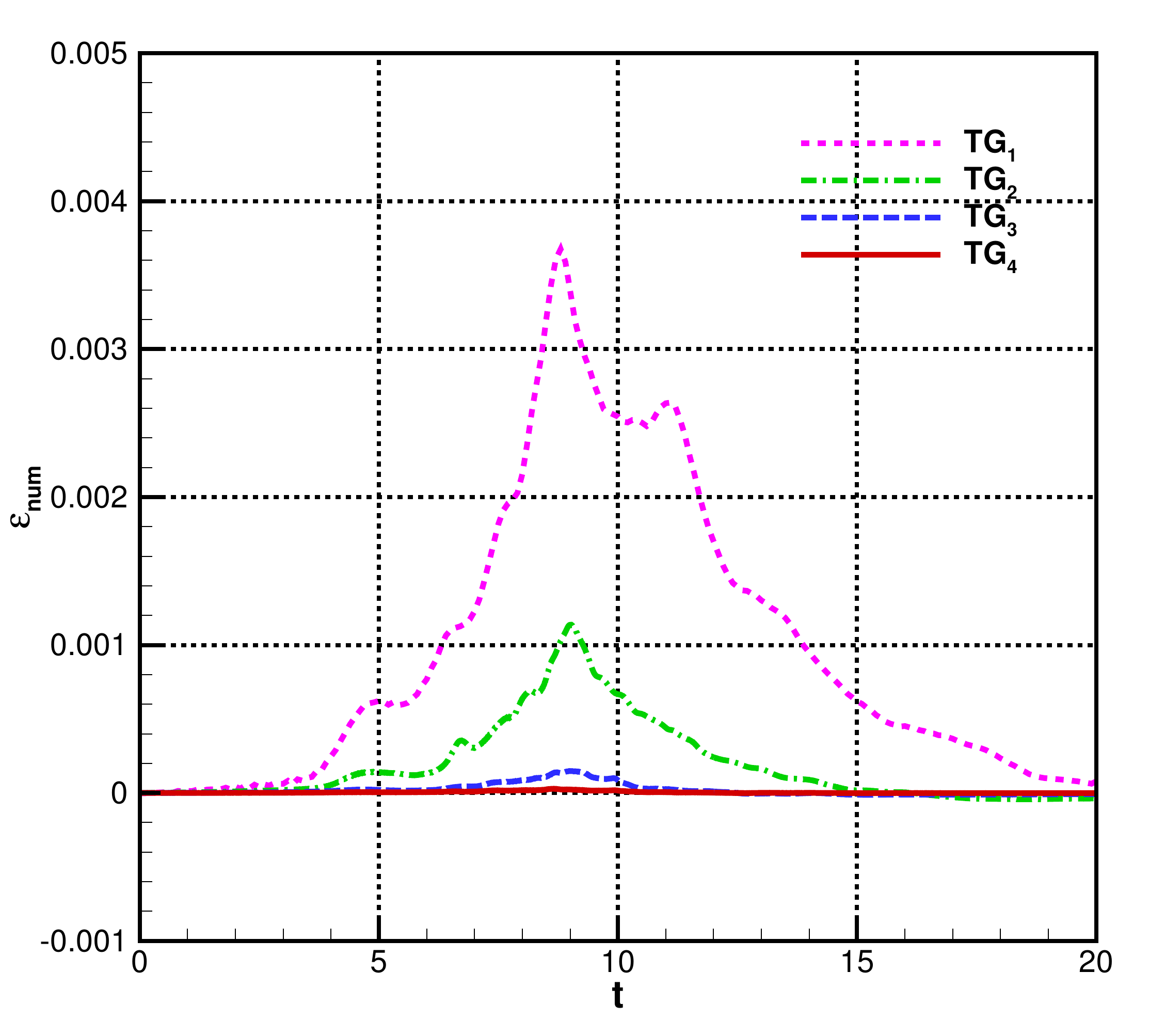}
\caption{\label{tg-vortex-3} Taylor-Green Vortex: the time
history of  $\varepsilon_1$, $\varepsilon_2$, $\varepsilon_3$ and
$\varepsilon_{num}$.}
\end{figure}

In the numerical simulation, the final dissipative behavior is determined by both
physical and numerical dissipation.
For the current study, the quantitative study of numerical dissipation is
presented as well, which is less reported in literatures. For the
compressible flow, the kinetic energy dissipation rate obtained
from the Navier-Stokes equations is the sum of three contributions, namely,
\begin{align*}
\varepsilon_1=&2\frac{\mu}{\rho_0}\frac{1}{\Omega}\int_\Omega
\boldsymbol{S}^d:\boldsymbol{S}^d\text{d}\Omega,\\
\varepsilon_2=&\frac{\mu_b}{\rho_0}\frac{1}{\Omega}\int_\Omega
(\nabla\cdot\boldsymbol{U})^2\text{d} \Omega,\\
\varepsilon_3=&-\frac{1}{\rho_0\Omega}\int_\Omega
p\nabla\cdot\boldsymbol{U} \text{d} \Omega,
\end{align*}
where $\boldsymbol{S}^d$ is the deviatoric part of the strain rate
tensor, $\mu_b$ is the bulk viscosity.  In current scheme, the
inherent bulk viscosity \cite{GKS-Xu1} reads
\begin{align*}
\mu_b=\frac{2N}{3(N+3)}\mu,
\end{align*}
where $N=2$ for the diatonic gas. The contributions to the
dissipation rate based on the compressible flow assumptions are
shown in Fig.\ref{tg-vortex-3}.
To eliminate the error from numerical discretization, all spatial derivatives are computed by sixth order central difference for three components of dissipation rate.
As excepted, the primary contribution $\varepsilon_1$  is almost identical to
$\varepsilon(\zeta)$ in current nearly incompressible simulation, and the bulk viscosity contribution
$\varepsilon_2$ and dilatation contribution $\varepsilon_3$ can be neglected.
It is noted that the magnitude of pressure dilation term $\varepsilon_3$ is on the same order as that of the reference solution \cite{Case-Debonis}.
With coarse mesh resolutions, the total dissipation rate
computed from $\varepsilon_1+\varepsilon_2+\varepsilon_3$ is significantly
lower than $\varepsilon(E_k)$.
Therefore, the  numerical dissipation can be quantitatively computed by
\begin{align*}
\varepsilon_{\text{num}}
=\varepsilon(E_k)-(\varepsilon_1+\varepsilon_2+\varepsilon_3).
\end{align*}
The time history of numerical dissipation is given in
Fig.\ref{tg-vortex-3} as well. With the refinement of grid, the
resolution of the vortical structures increases  and the effect of
the filtering decreases, which reduces numerical dissipation. The fact
that $\varepsilon(E_k)$ is well predicted at all grid levels
indicates that the physical and numerical dissipation work together consistently in the calculation
to get the final "physical" result. In other words, the filtering due to the coarse mesh correctly mimics
the physical dissipation from the unresolved scale dynamics.
This validates the usage of high-order numerical methods for the implicit
large-eddy simulations (iLES) \cite{iles}.
While the iLES for complex turbulent flows is still under debate,
the current quantitative analysis of numerical dissipation gives the specific hints on this issue.

\subsection{Efficiency test of parallel computation}
\begin{figure}[!h]
\centering
\includegraphics[width=0.5\textwidth]{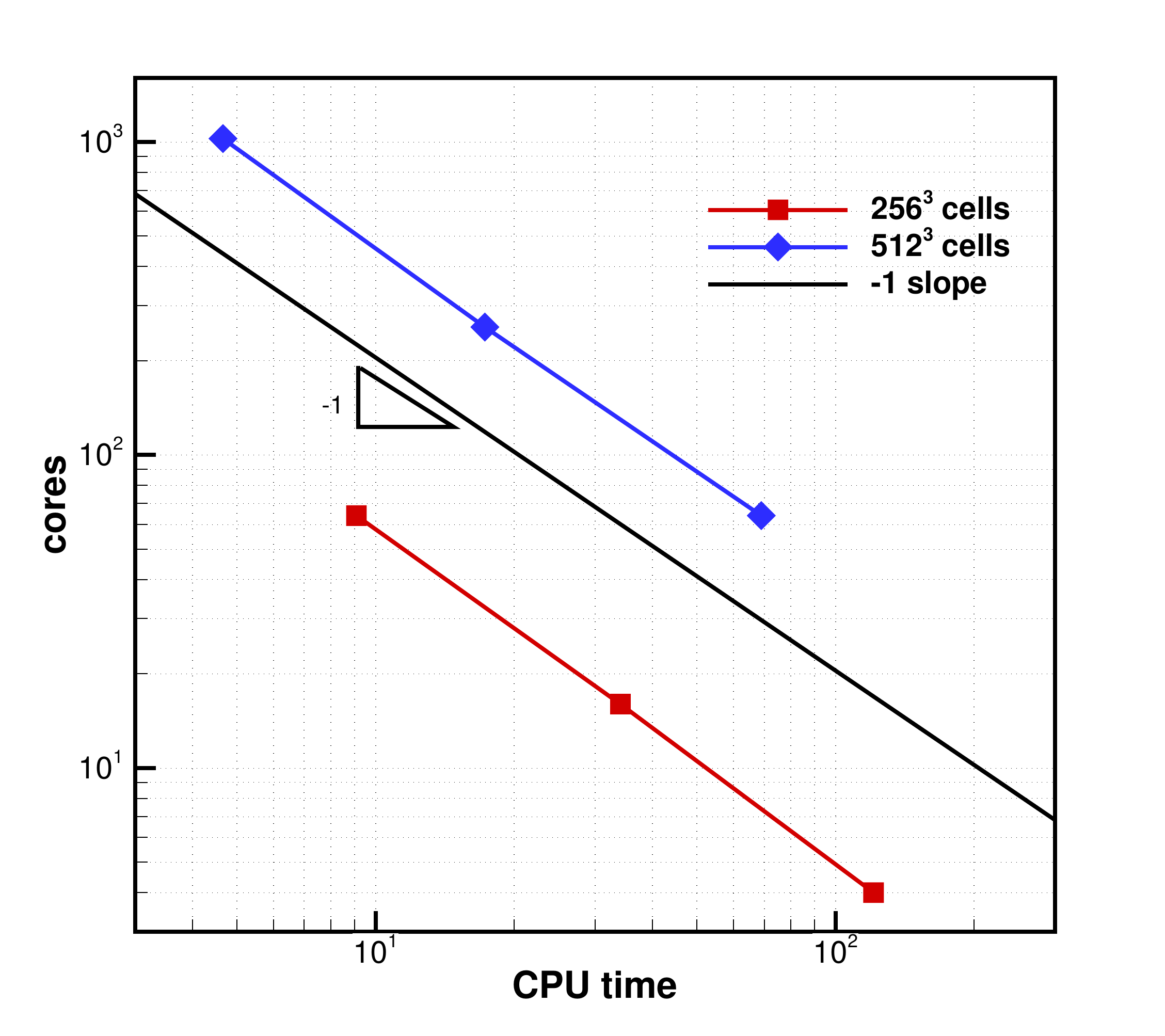}
\caption{\label{time_speedup}  Efficiency test of parallel strategy:  log-log plot for $n$ and $T_n$.}
\end{figure}

In this section, the efficiency of parallel computation is tested in
the above Taylor-Green vortex problem. To give the performance of parallel
computation, the speedup is defined as
\begin{align*}
S_{n}=\frac{T_n}{T_{n_{\text{ref}}}},
\end{align*}
where $T_n$ is the execution time with $n$ cores and
$T_{n_{\text{ref}}}$ is the execution time on a reference number of
processors. The ideal speedup of parallel computations would be
equal to $n/n_{\text{ref}}$. With the log-log plot for $n$ and
$T_n$, an ideal scalability would follow $-1$ slope. However, this
efficiency is not possible due the communication delay among
the computational cores and the idle time of computational nodes
associated with load balancing. The scalability of our MPI code is
examined by measuring the wall clock time against the number of
processors. The detailed performance of MPI parallel computing is
given in Table.\ref{time_table1} and a log-log plot is also given in
Fig.\ref{time_speedup},  where $4$, $16$ and $64$ cores for  $256^3$
cells, and $64$, $256$, and $1024$ cores are used for $512^3$ cells.
Total $30$ steps are computed for each case, and CPU time is the
averaged time for each step. The code was run on the TianHe-II, and
the node details are presented in Table.\ref{time_tainhe}. Due to
the explicit formulation of HGKS, our MPI code scales properly with
the number of processors used.
It is indicated that the data communication crossing nodes costs a little time and the computation for flow field is the dominant one.

\begin{table}[!h]
    \centering
    \begin{tabular}{c|c|c|c|c|c}
        \hline
        \hline
        Grid size &Cores & CPU time (s/step) &Grid size &Cores & CPU time (s/step)  \\
        \hline
        $256^3$ &4 &  121.01   &$512^3$ &64 & 69.00 \\
        \hline
        $256^3$ &16 & 34.08    &$512^3$ &256 & 17.29  \\
        \hline
        $256^3$ &64 & 9.09     &$512^3$ &1024 & 4.66 \\
        \hline
        \hline
    \end{tabular}
    \caption{\label{time_table1} Efficiency test of parallel strategy: detailed CPU time against
number of core.}
\end{table}

\begin{table}[!h]
    \centering
    \begin{tabular}{c|c|c|c}
        \hline
        \hline
        System        &Node type                         & Cache     &Host channel adapter\\
        \hline
        TianHe-II     &Intel Xeon E5-2692 (2.2GHz/core)   & 30MB     &InfiniBand \\
        \hline
        Pleiades      &Intel Xeon E5-2670 (2.6GHz/core)   & 20MB     &InfiniBand  \\
        \hline
        \hline
    \end{tabular}
    \caption{\label{time_tainhe} Efficiency test of parallel strategy:  node details for TianHe-II and Pleiades supercomputer systems.}
\end{table}

For most DNS, the high-order finite difference method is widely
used, and the finite volume scheme is rarely applied  due to the
complicated formulation. Because of the procedure of
multidimensional spatial reconstruction and quadrature of numerical
fluxes at cell interface, the finite volume scheme is considered to
be less efficient than the finite difference method. For the
high-order gas-kinetic scheme, the gas-kinetic flux solver
Eq.\eqref{flux} considered to be even more complicated than Riemann solvers
\cite{Riemann-appro}, which is usually used in the classical finite
volume scheme. In this case, the comparison of total computational
cost with the finite difference method is also given. The time step
and total computational costs of HGKS are presented in
Table.\ref{time_table2}, in which the current HGKS is running on the
TianHe-II supercomputer system. As reference, the efficiency of high-order
finite difference method \cite{Case-Debonis} is given in
Table.\ref{time_table3}, in which the cases were run on the NASA
Pleaides high performance computing system. The node details are
compared in Table.\ref{time_tainhe}, and the total computational costs
of HGKS is around $1.4$ times higher than the finite difference
method. In addition, considering the processor speed, the HGKS is
around $1.2$ times higher than the finite difference method. Taken
the robustness of HGKS into account, such computational cost is
comparable and affordable.

\begin{table}[!h]
    \centering
    \begin{tabular}{c|c|c|c|c|c}
        \hline
        \hline
        Case  &   Grid size & Time step            & Cores & Hours   & Computational costs\\
        \hline
        $TG_1$ & $128^3$  & $1.785\times10^{-3}$  & 16    & 13.3    & 213 core hours\\
        \hline
        $TG_2$ & $256^3$  & $8.925\times10^{-4}$  & 256   & 13.5    & 3456 core hours\\
        \hline
        $TG_3$ & $512^3$  &$4.462\times10^{-4}$   & 1024  & 66      & 67584 core hours\\
        \hline
        $TG_4$ &  $1024^3$ &$2.789\times10^{-4}$   & 1024  & 730     & 747062 core hours\\
        \hline
        \hline
    \end{tabular}
    \caption{\label{time_table2} Efficiency test of parallel strategy: detailed computational parameters for HGKS.}
\end{table}

\begin{table}[!h]
    \centering
    \begin{tabular}{c|c|c|c|c}
        \hline
        \hline
        Grid size & Time step             & Cores & Hours  & Computational costs\\
        \hline
        $256^3$   & $8.463\times10^{-4}$  & 64    & 40     & 2560 core hours\\
        \hline
        $512^3$   &$4.231\times10^{-4}$   & 368   & 130    & 47840 core hours\\
        \hline
        \hline
    \end{tabular}
    \caption{\label{time_table3} Efficiency test of parallel strategy: detailed computational parameters for BB13 \cite{Case-Debonis}.}
\end{table}

\subsection{Turbulent channel flow}
\begin{figure}[!h]
\centering
\includegraphics[width=0.65\textwidth]{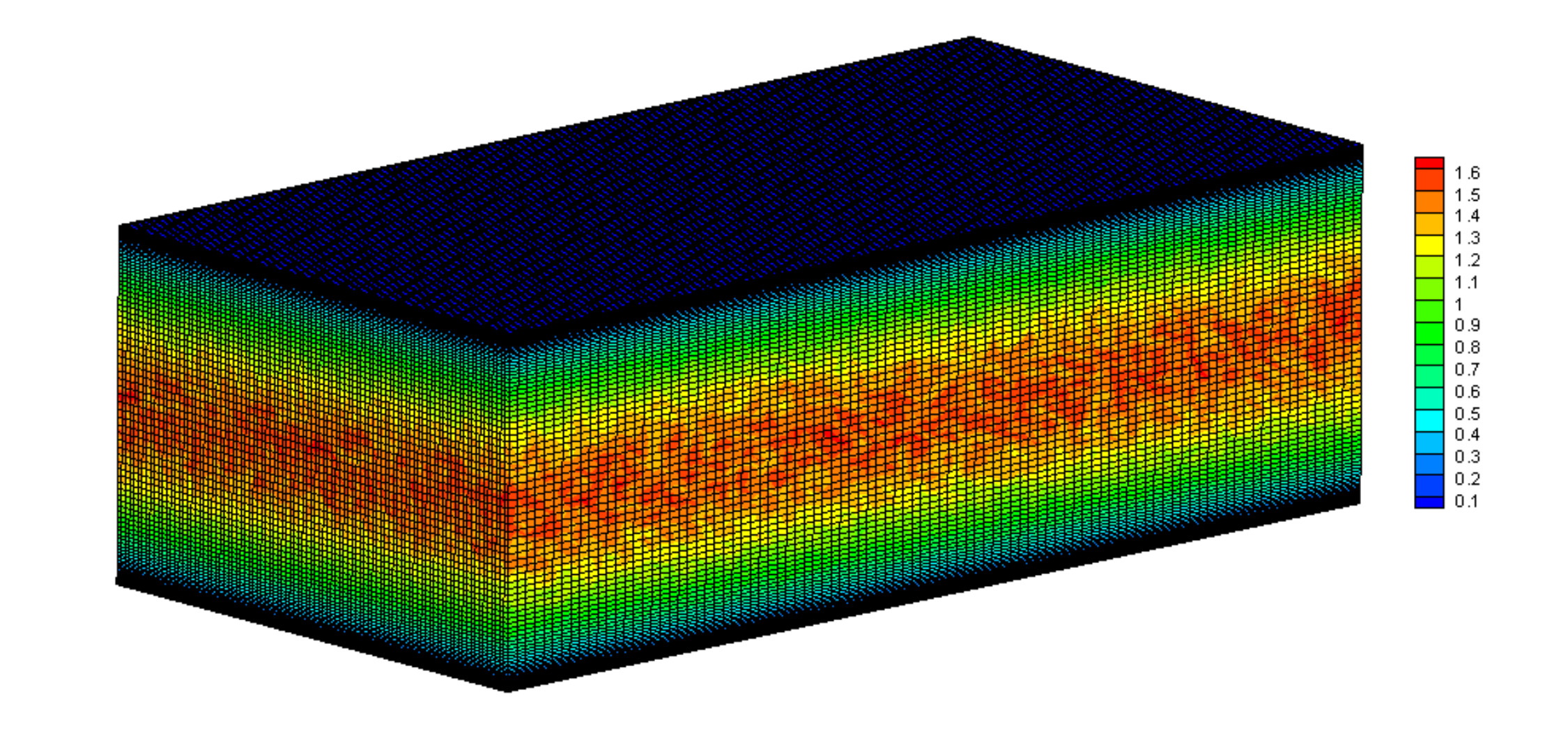}
\caption{\label{channel_mesh} Turbulent channel flow: the mesh and initial streamwise velocity distributions for case $G_2$.}
\end{figure}

Considering the simplicity of geometry and boundary conditions, the
turbulent channel flows have been studied to understand the
mechanism of wall-bounded turbulent flows.  A large number of
computational studies of turbulent channel flows have been carried
out  \cite{kim1987turbulence,hoyas2006scaling,lee2015direct}.
In the current study, the
turbulent channel flow with friction Reynolds number $Re_\tau=180$
is tested. In the computation, the physical domain is
$(x,y,z)\in[0,2\pi]\times[-1,1]\times[0,\pi]$ and the computational domain takes
$(\xi,\eta,\zeta)\in[0,2\pi]\times[0,3\pi]\times[0,\pi]$. The coordinate
transformation is given by
\begin{align*}
\begin{cases}
\displaystyle x=\xi,\\
\displaystyle y=\tanh(b_g(\frac{\eta}{1.5\pi}-1))/\tanh(b_g),\\
\displaystyle z=\zeta,
\end{cases}
\end{align*}
where $b_g=2$.  The periodic
boundary conditions are used in streamwise $x$-direction and spanwise $z$-directions, and the non-slip and isothermal
boundary conditions are used in vertical $y$-direction. The fluid is initiated with $\rho=1, Ma=0.1$ and the
initial streamwise velocity profile is given by the perturbed Poiseuille flow
profile
\begin{align*}
U(y)&=1.5(1-y^{2}) + \text{white noise}.
\end{align*}
 White noise is added with $10\%$ amplitude of
local streamwise velocity.
With the unit averaged streamwise velocity, the initial pressure
can be given. The friction Reynolds number is
defined as
\begin{align*}
Re_\tau=\rho u_\tau H/\mu,
\end{align*}
where $H = 1$ is the half height of the channel and the frictional velocity $u_\tau$ is given by
\begin{align*}
u_\tau = \sqrt{\frac{\tau_{wall}}{ \rho}}, ~\tau_{wall}=\frac{\partial U}{\partial y}\big|_{wall}.
\end{align*}
For the channel flow, the logarithmic formulation is given by
\begin{align}\label{uplus}
U^+ = \frac{1}{\kappa}\ln Y^+ + B,
\end{align}
where von Karman constant $\kappa=0.40$ and $B=5.5$ for the low Reynolds number
turbulent channel flow \cite{kim1987turbulence}.
The plus unit and plus velocity are defined as
\begin{align*}
Y^+ = \rho u_\tau y / \mu, ~U^+ = U/u_{\tau}.
\end{align*}
Therefore, the plus velocity  $U_c^{+} =18.4823$ at center line of
the channel according to Eq.\eqref{uplus}, where $Y_c^{+} =
180$ at center line.
The frictional velocity is determined by
$u_{\tau} = U_c/U_c^{+} = 0.0541$, where $U_c=1$ is the centreline
line velocity.  In this computation, the cases $G_1$ and $G_2$ are tested, where $96^3$ and $128^3$ cells are distributed uniformly in computational space.
$256$ cores and $1024$ cores are used to simulate the $G_1$ and $G_2$, respectively.
The details of mesh are given in Table.\ref{channel_grids},  where $\Delta y^{+}_{min}$ and $\Delta y^{+}_{max}$ are  the minimum and maximum grid space in the $y$-direction.
To resolve the viscous layer, there are $11$ layers for $G_1$ and $15$
layers for $G_2$ within $Y^+$ less than $10$, respectively.
As reference, the mesh and initial streamwise velocity in the physical
domain for $G_1$ are given in Fig.\ref{channel_mesh}.
\begin{table}[!h]
    \centering
    \begin{tabular}{c|c|c|c|c}
        \hline \hline
        Case    &Grid size  &$\Delta y^{+}_{min}$/$\Delta y^{+}_{max}$   &$\Delta x^{+}$   &$\Delta z^{+} $\\
        \hline
        $G_1$   &$96^3$  &0.29/7.77   &11.77   &5.89  \\
        \hline
        $G_2$   &$128^3$ &0.21/5.83   &8.83    &4.42 \\
        \hline \hline
    \end{tabular}
    \caption{\label{channel_grids} Turbulent channel flow: different sets of grids for $Re_{\tau} = 180$ turbulent channel flow.}
\end{table}

\begin{figure}[!h]
    \centering
    \includegraphics[width=0.475\textwidth]{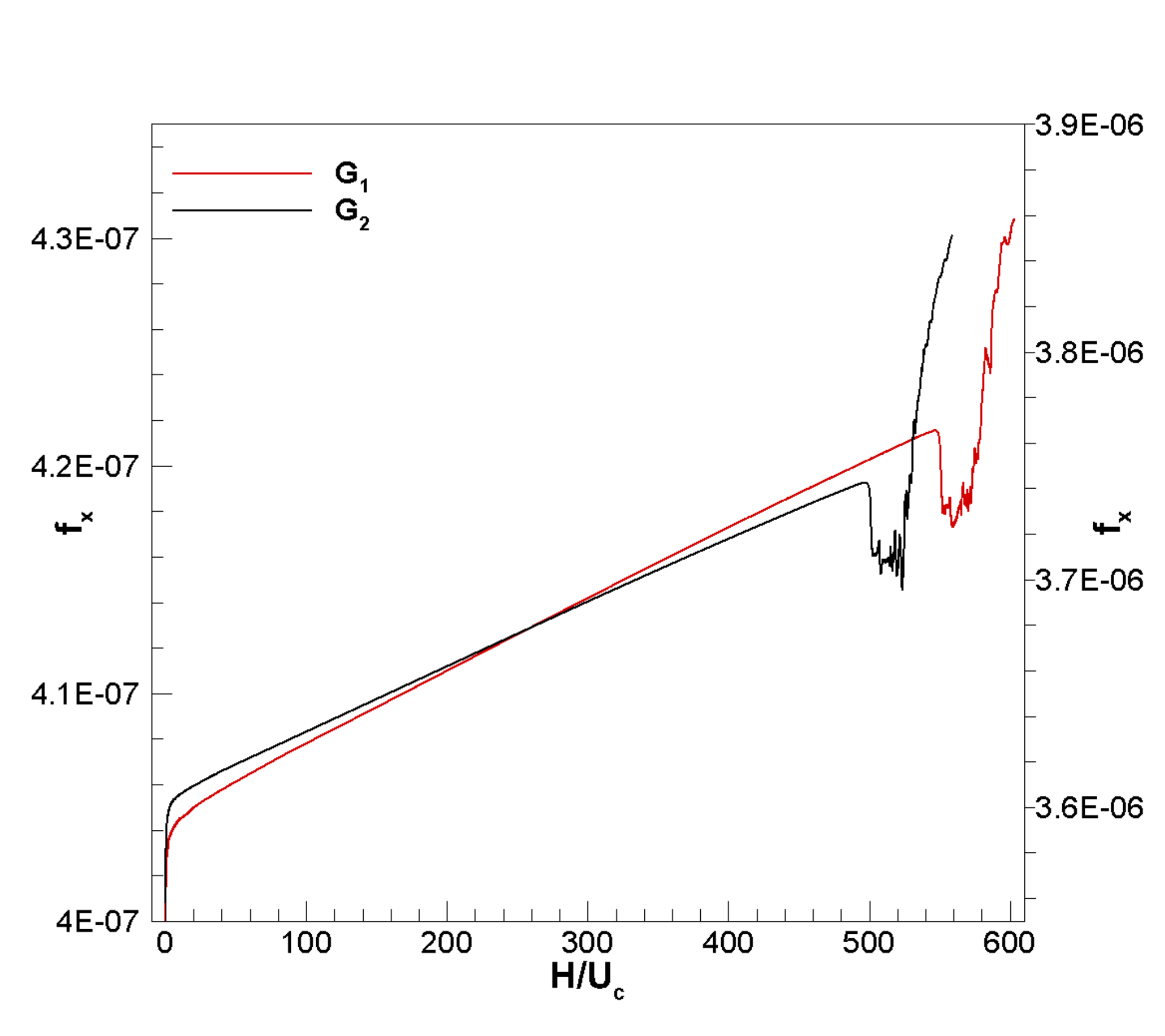}
    \includegraphics[width=0.475\textwidth]{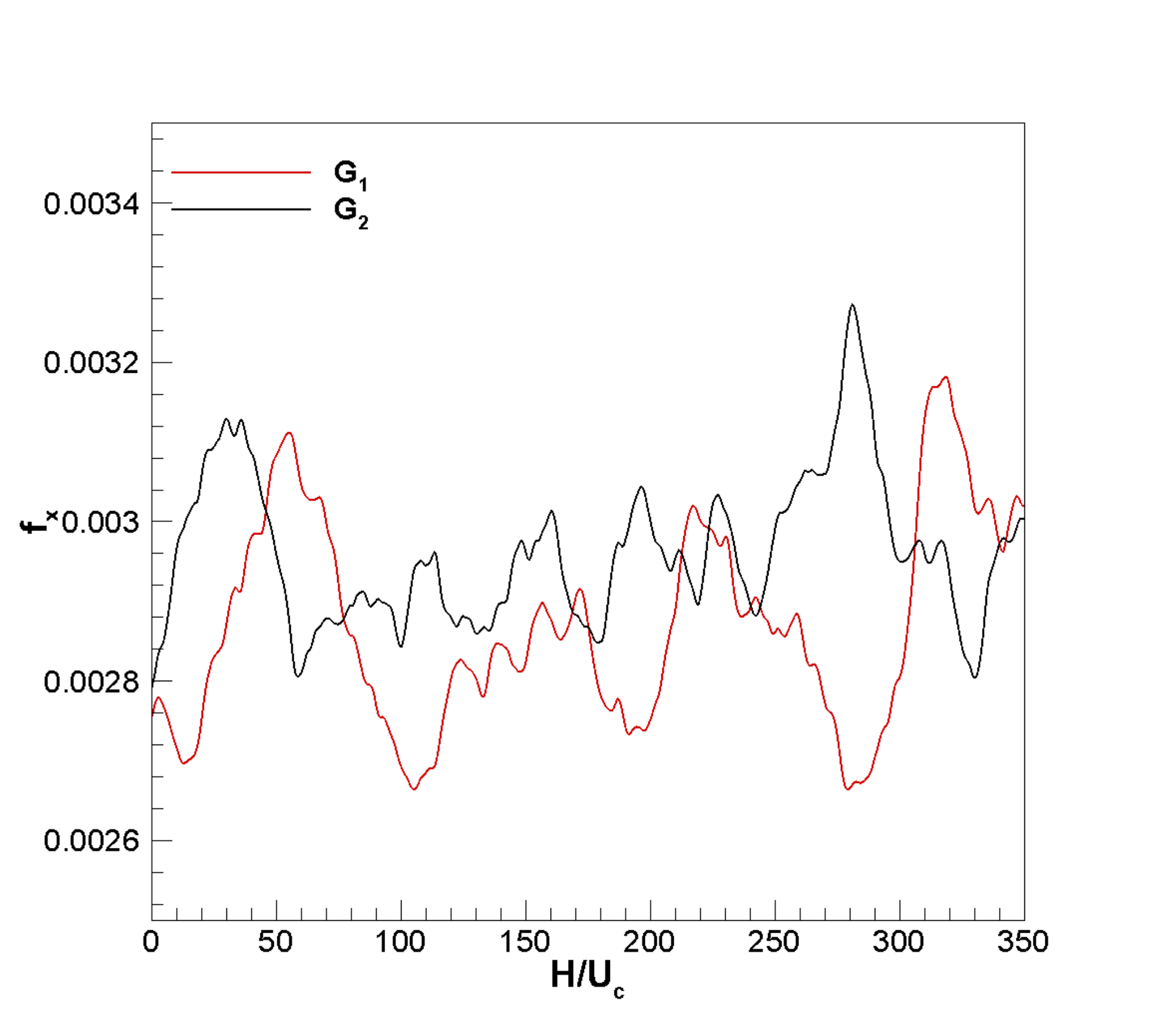}
    \caption{\label{channel_force} Turbulent channel flow: time evolution of the external force for case $G_1$ and $G_2$.}
    \centering
    \includegraphics[width=0.475\textwidth]{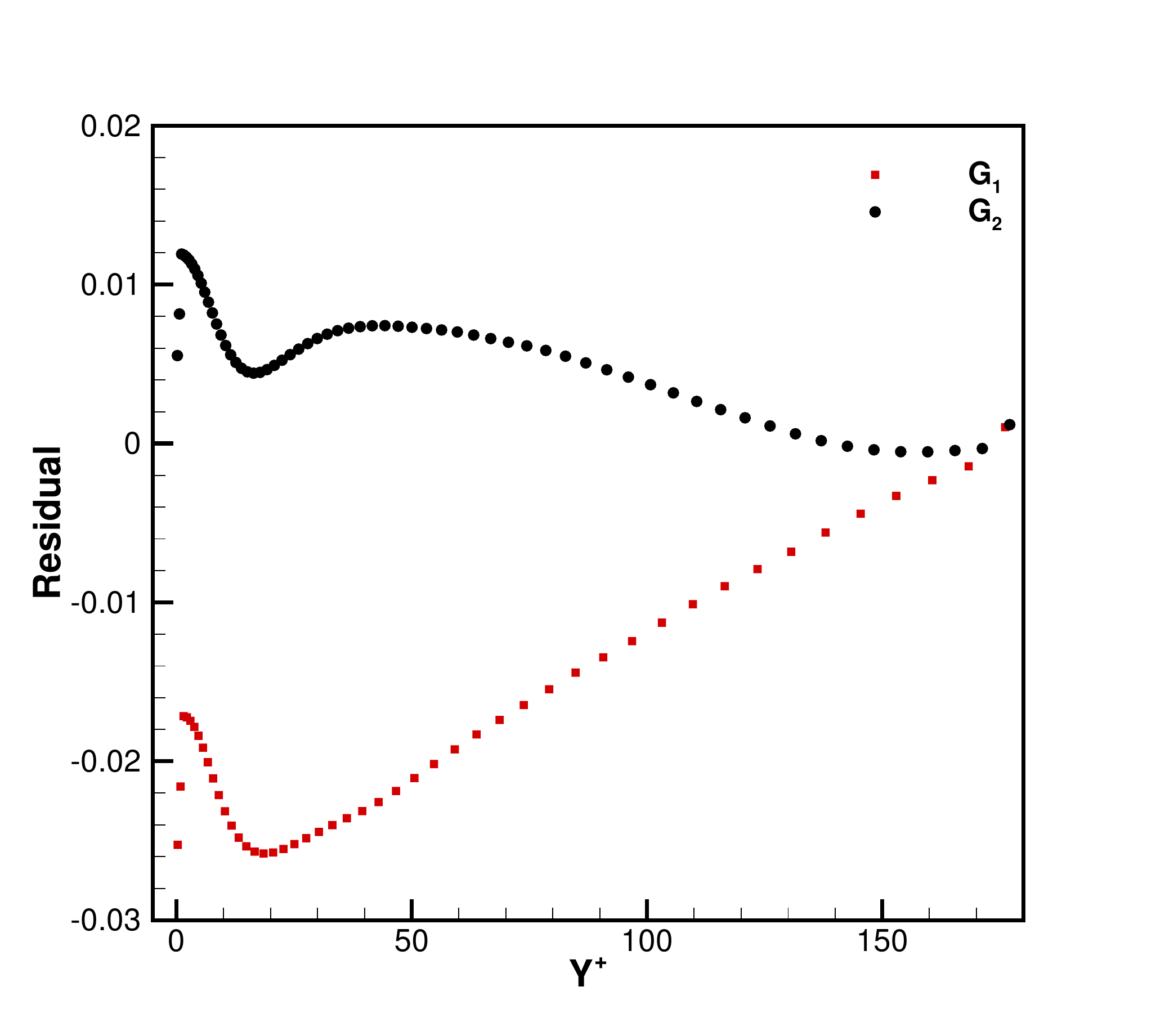}
    \caption{\label{channel_stress_residual} Turbulent channel flow: distribution of stress residual for case $G_1$ and $G_2$.}
\end{figure}

To excite channel flow from laminar to turbulence, an external force
is exerted in the streamwise direction.
According to the viscous layer $U^+ = Y^+$, a fixed nondimensional external force can be approximated by balance
of forces
\begin{align*}
f_x=\tau_{wall} /H= 2.93\times 10^{-3}.
\end{align*}
Before transition, the external force $f_x\cdot \Delta t$ and $10 f_x\cdot
\Delta t$ are used for $G_1$ and $G_2$, respectively. $\Delta t$ is the time step.
After transition, the constant moment flux is used to determine the external force.
According to the experiment and previous work \cite{kim1987turbulence,DUGKS}, the constant bulk volume is recommended
to be set as
\begin{align*}
\displaystyle\iiint_\Omega(\rho U)^{n+1}_{ijk}\text{d}\Omega = \displaystyle\iiint_\Omega(\rho U_b)\text{d}\Omega,
\end{align*}
where $U_c/U_b = 1.16$ is chosen based on previous DNS \cite{kim1987turbulence}.
The conservative variables updated by the two-stage method
Eq.\eqref{two-stage} are denoted as $\widetilde{Q}^{n+1}$ and the
conservative variables with external force is $Q^{n+1}$. With the
external force, the equation for momentum in streamwise direction
and energy can be written as
\begin{align*}
\frac{\partial \rho U}{\partial t}&=\mathcal {L}_{\rho U}+\rho f_x,\\
\frac{\partial \rho E}{\partial t}&=\mathcal {L}_{\rho E}+\rho U
f_x,
\end{align*}
where $\mathcal {L}_{\rho U}, \mathcal {L}_{\rho E}$ are the
operator for spatial derivative of momentum and energy fluxes and
$f_x$ is the external force, which can be given as follows
\begin{align*}
f_x=\frac{1}{\Delta t}\cdot\frac{\displaystyle\iiint_\Omega(\rho
    U)^{n+1}_{ijk}\text{d}\Omega-\iiint_\Omega(\widetilde{\rho
        U})^{n+1}_{ijk}\text{d}\Omega}{\displaystyle\frac{1}{2}\iiint_\Omega(\rho^n_{ijk}+\widetilde{\rho}^{n+1}_{ijk})\text{d}\Omega}.
\end{align*}
and the equation for momentum in streamwise direction and energy can
be updated
\begin{align*}
(\rho U)_{ijk}^{n+1}&=(\widetilde{\rho U})_{ijk}^{n+1}+\frac{1}{2}\Delta t(\rho^{n+1}_{ijk}+\rho^{n}_{ijk})f_x,\\
(\rho E)_{ijk}^{n+1}&=(\widetilde{\rho
    E})_{ijk}^{n+1}+\frac{1}{2}\Delta t((\rho U)^{n+1}_{ijk}+(\rho
U)^{n}_{ijk})f_x.
\end{align*}
Therefore, the momentum flux over the whole domain keeps constant in the computation.
The external force before the transition and after transition
with grid $G_1$ and $G_2$ are presented in Fig.\ref{channel_force}.
After over $500$ characteristic periodic time as
$500H/U_c$, it can be seen that the initial laminar flow-fields transit
to turbulence. Then, the external force based on the constant moment  flux are used.
The total stress is used to test whether the
simulated turbulence is  statistically stationary
\cite{tennekes1972first, lee2015direct}. In a statistically
stationary turbulent channel, the total stress, which is the sum of
Reynolds stress and mean viscous stress, is linear because of
momentum conservation
\begin{align*}
\frac{\text{d} U^+}{\text{d} Y^+} - \left\langle UV \right\rangle^+ \approx 1 - \frac{Y^+}{Re_{\tau}}.
\end{align*}
When the residual of total stress converging, the $350$ periodic
time as $350H/U_c$ is used, which is comparable to that in the reference paper
\cite{vreman2004eddy,hoyas2006scaling}. As shown in
Fig.\ref{channel_stress_residual}, the residual is less than $2.6\%$
for case $G_1$ and $1.2\%$ for case $G_2$.

\begin{table}[!h]
    \centering
    \begin{tabular}{c|c|c|c|c}
        \hline \hline
        Case     &$Re_{\tau}$ & $Re_{c}$ & $Re_{b}$  &$C_f$ \\
        \hline
        $G_1$    & 176.03   &3297.78   &5720.09 & $7.58 \times 10^{-3}$ \\
        \hline
        $G_2$    & 179.21   &3319.22   &5730.11 & $7.82 \times 10^{-3}$ \\
        \hline \hline
    \end{tabular}
    \caption{\label{table_grid_mean} Turbulent channel flow: mean flow variables for case $G_1$ and $G_2$.}
\end{table}
\begin{figure}[!h]
    \centering
    \includegraphics[width=0.475\textwidth]{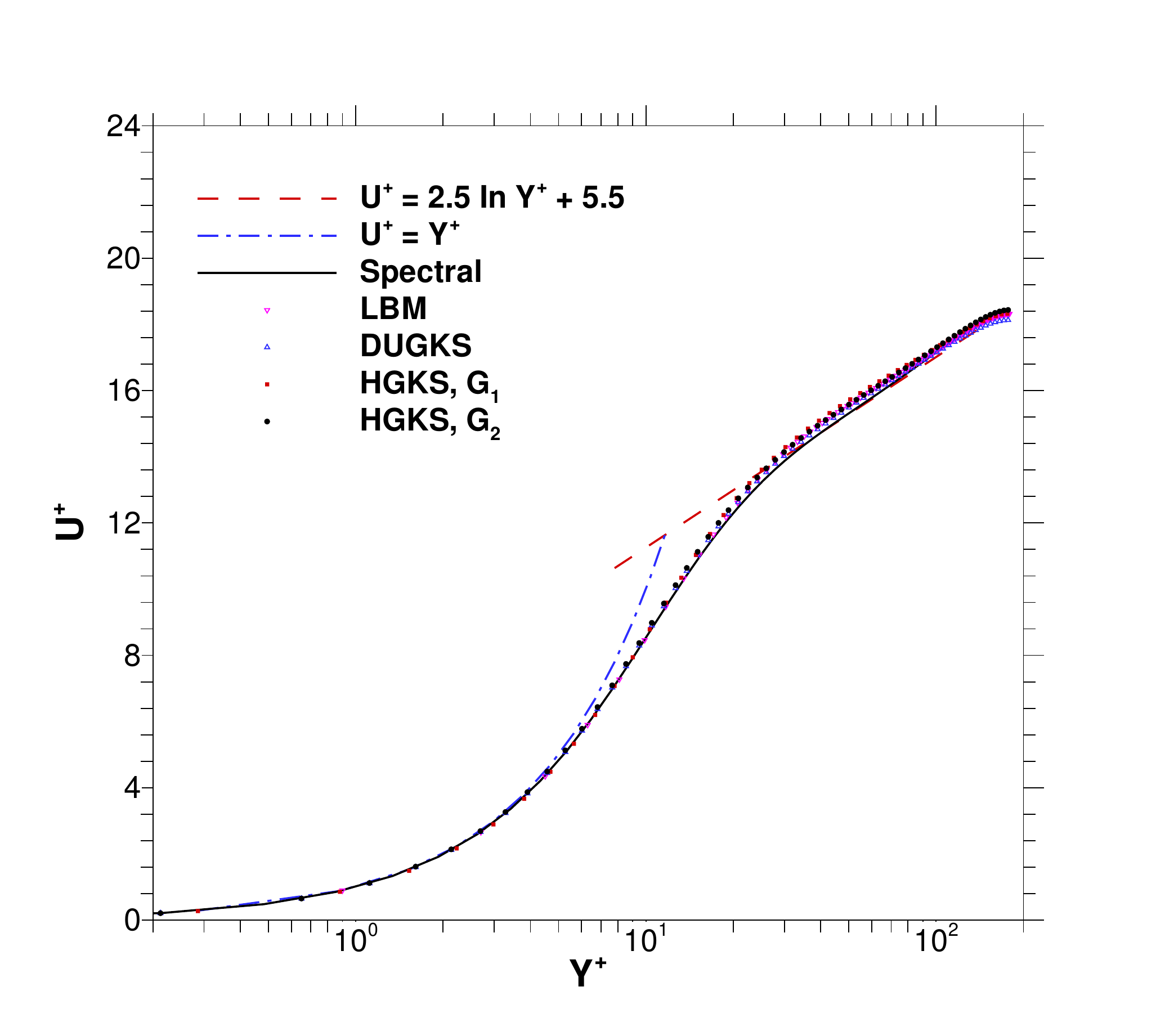}
    \includegraphics[width=0.475\textwidth]{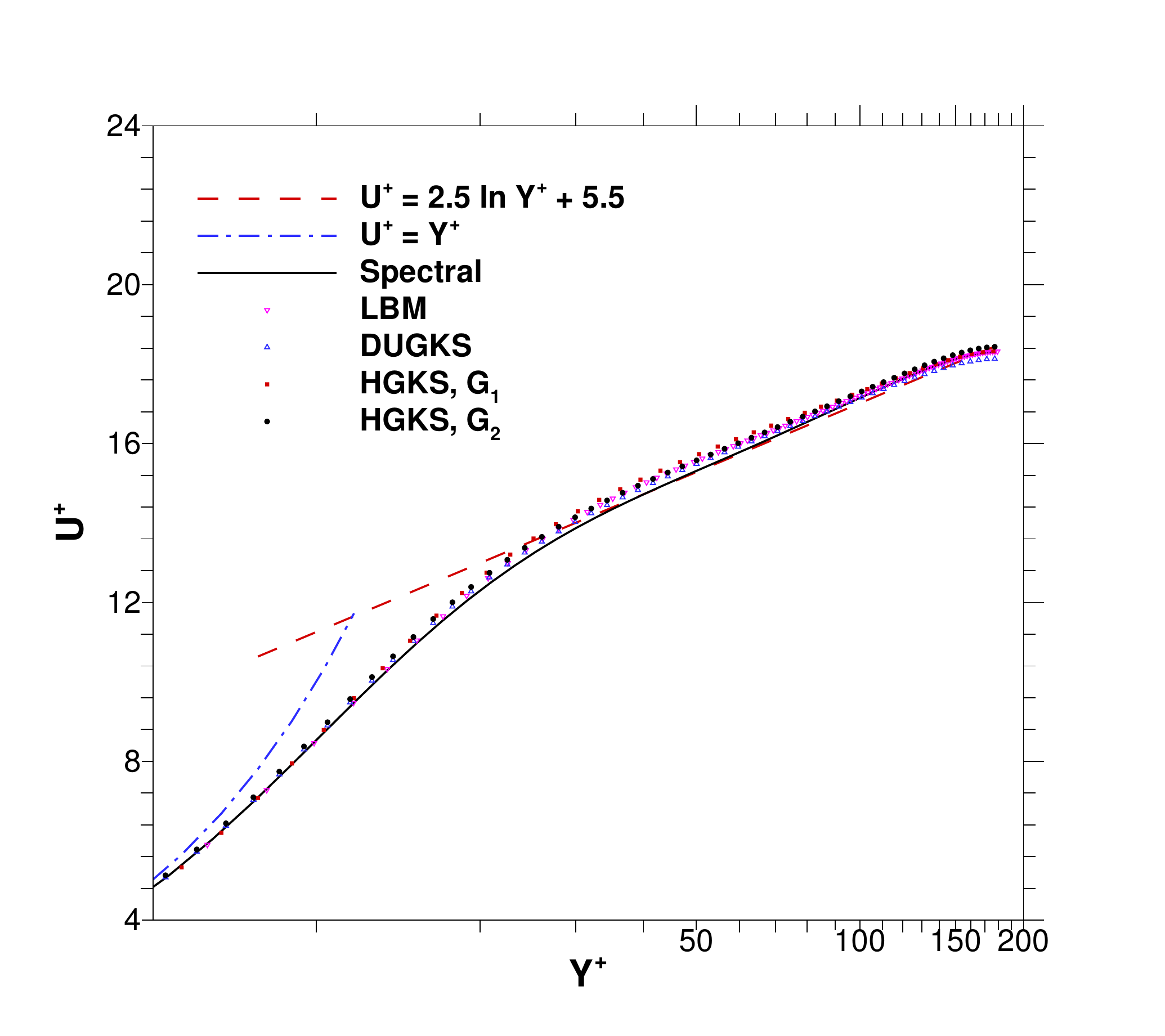}
    \caption{\label{channel_velocity} Turbulent channel flow: mean flow velocity profiles and the local enlargement.}
\end{figure}
The mean flow variables with different sets of grids are
presented in Table.\ref{table_grid_mean}.  $Re_{\tau}$ is the averaged friction Reynolds number,
$U_c$ is the averaged mean centerline velocity, $U_b$ is the averaged mean bulk
velocity. $Re_c = U_c H/\nu$, $Re_b = U_b 2H/\nu$, $C_f =
\tau_{wall}/(\rho U_b^2/2)$, and $C_f$ is the skin friction
coefficients. The result on $G_2$ is much closer with Dean's
suggested correlation of $C_f = 0.073Re_b^{-0.25} = 8.39 \times
10^{-3}$ \cite{dean1978reynolds}.
The averaged velocity profiles
with grid $G_1$ and $G_2$ are presented in Fig.\ref{channel_velocity}.
The first DNS of fully developed
incompressible turbulent channel flow was performed  by the spectral
method with $129 \times 192 \times 160$ grids \cite{kim1987turbulence}. As the most popular mesoscopic
methods for simulating nearly incompressible flows, the numerical
results of the LBM with $200 \times 400 \times 200$ grids
and DUGKS with $128^3$ grids are also presented \cite{DUGKS}, in which
the physical accuracy has been demonstrated by comparing with the
Navier-Stokes based spectral methods.
The mean flow velocity with a log-linear plot and a local enlargement are given
in Fig.\ref{channel_velocity}, where the HGKS result is in
reasonable agreement with the spectral results,  LBM and DUGKS results.

\begin{figure}[!h]
	\centering
	\includegraphics[width=0.475\textwidth]{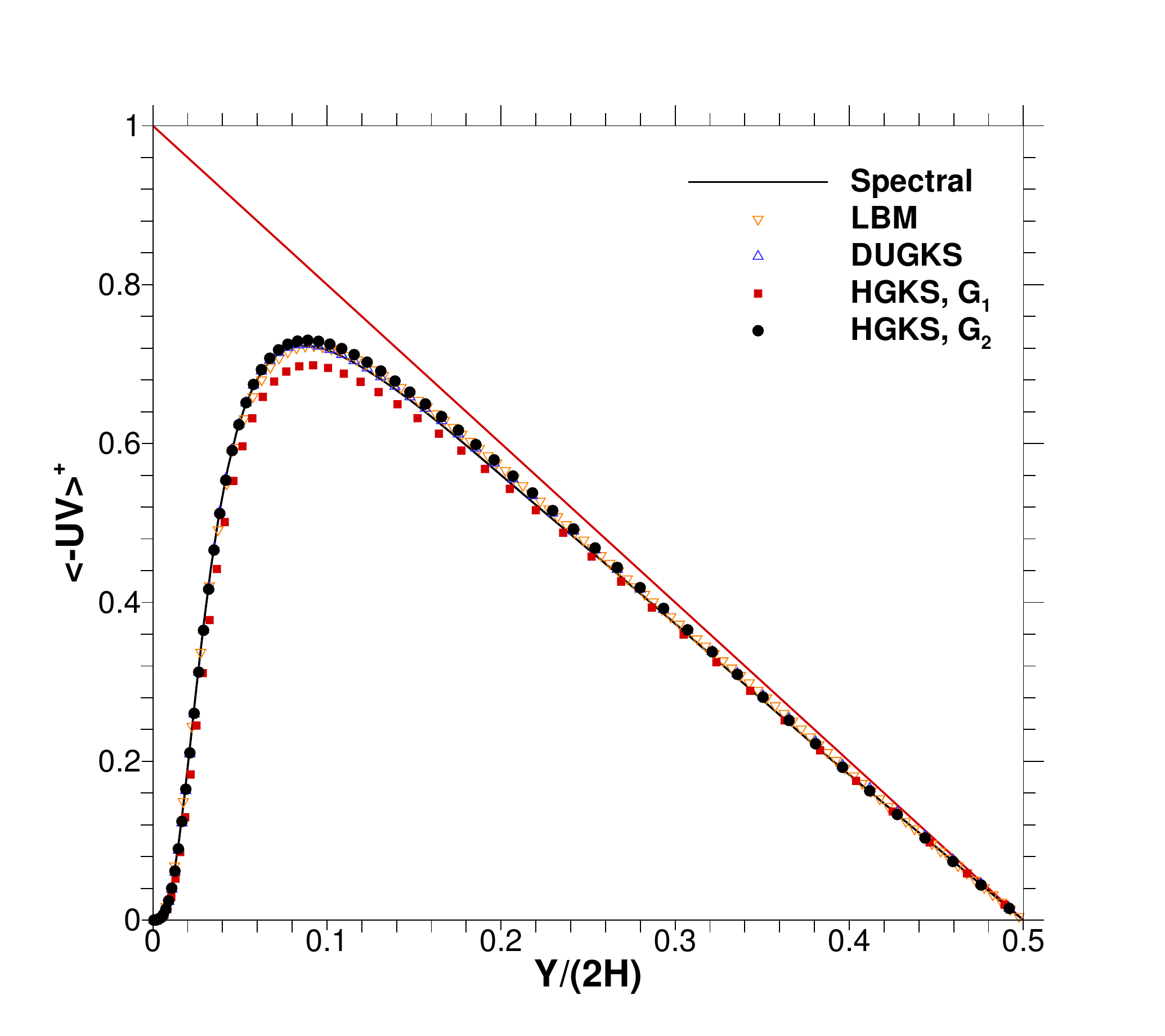}
	\includegraphics[width=0.475\textwidth]{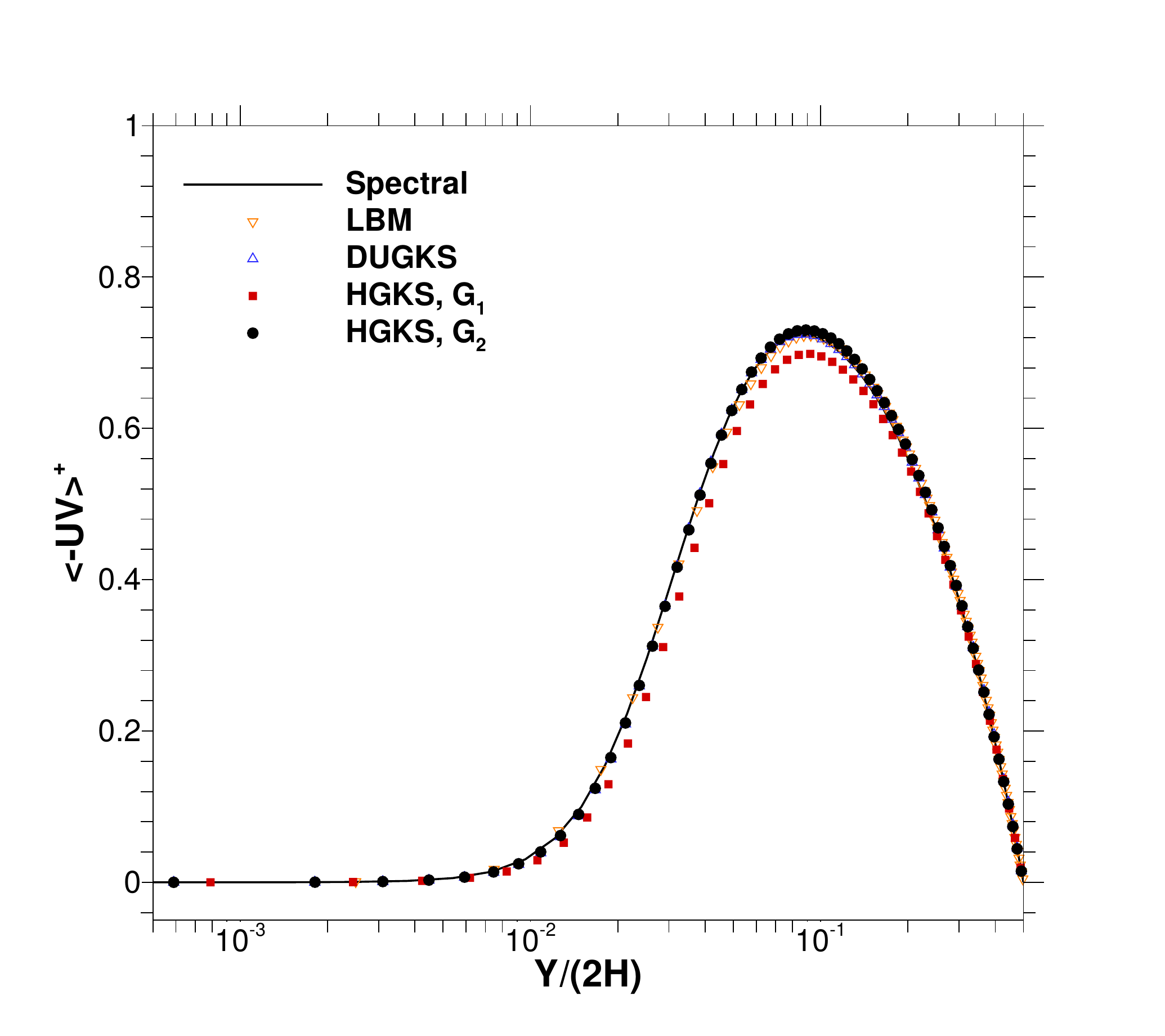}
	\caption{\label{channel_velocity-2} Turbulent channel flow: Reynolds stress profiles in linear-linear and log-linear plots.}
\end{figure}
\begin{figure}[!h]
	\centering
	\includegraphics[width=0.45\textwidth]{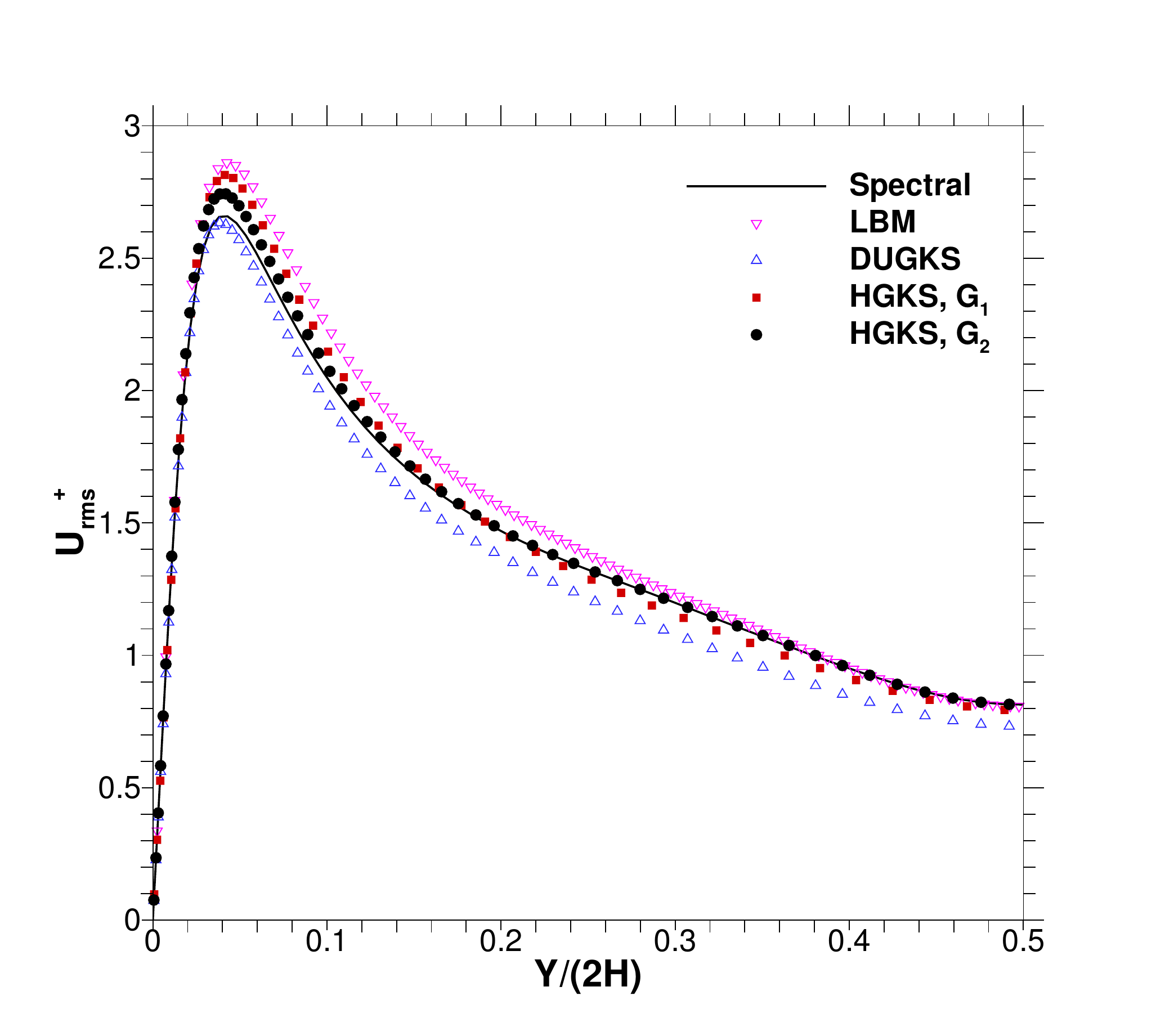}
	\includegraphics[width=0.45\textwidth]{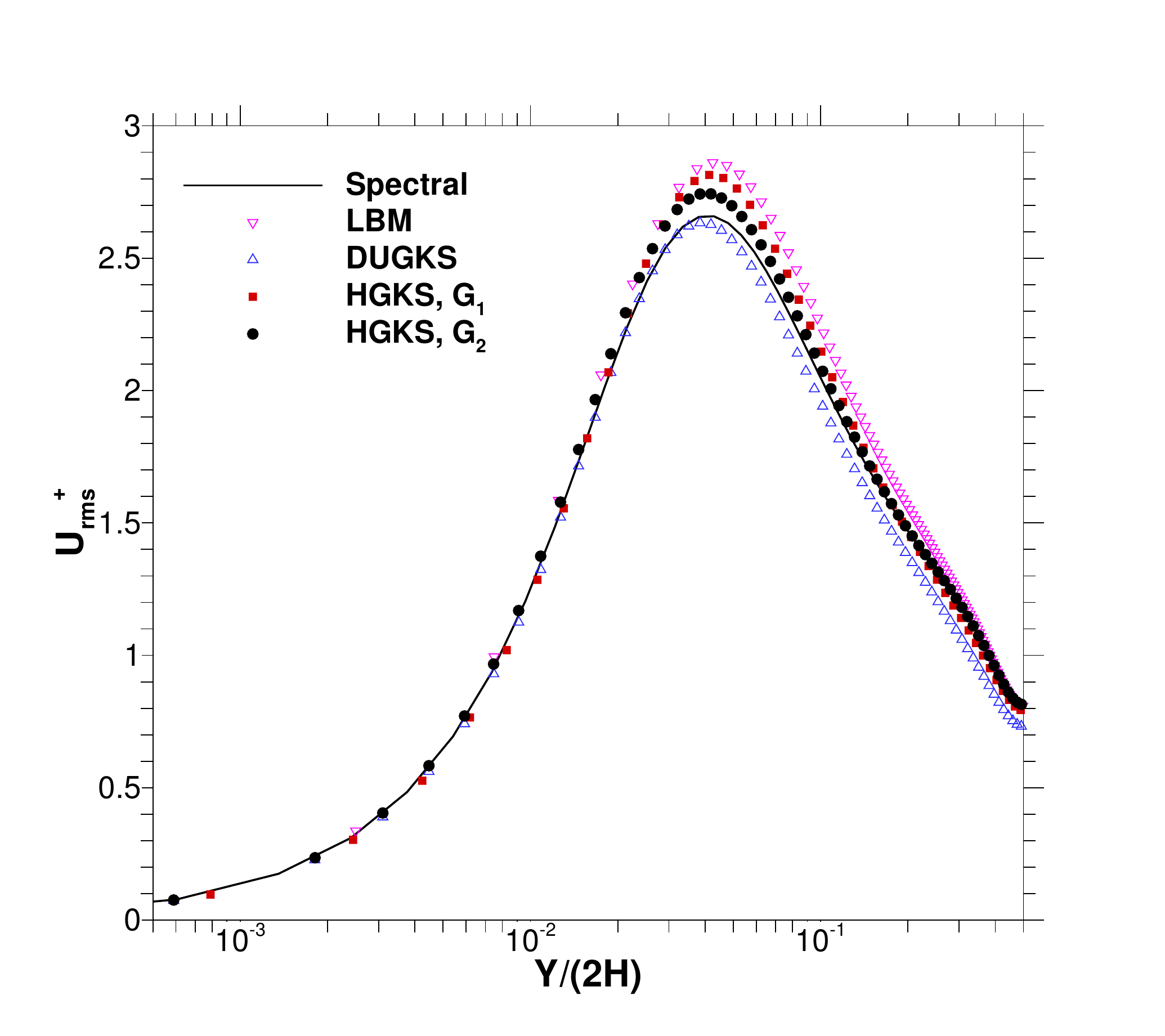}
	\includegraphics[width=0.45\textwidth]{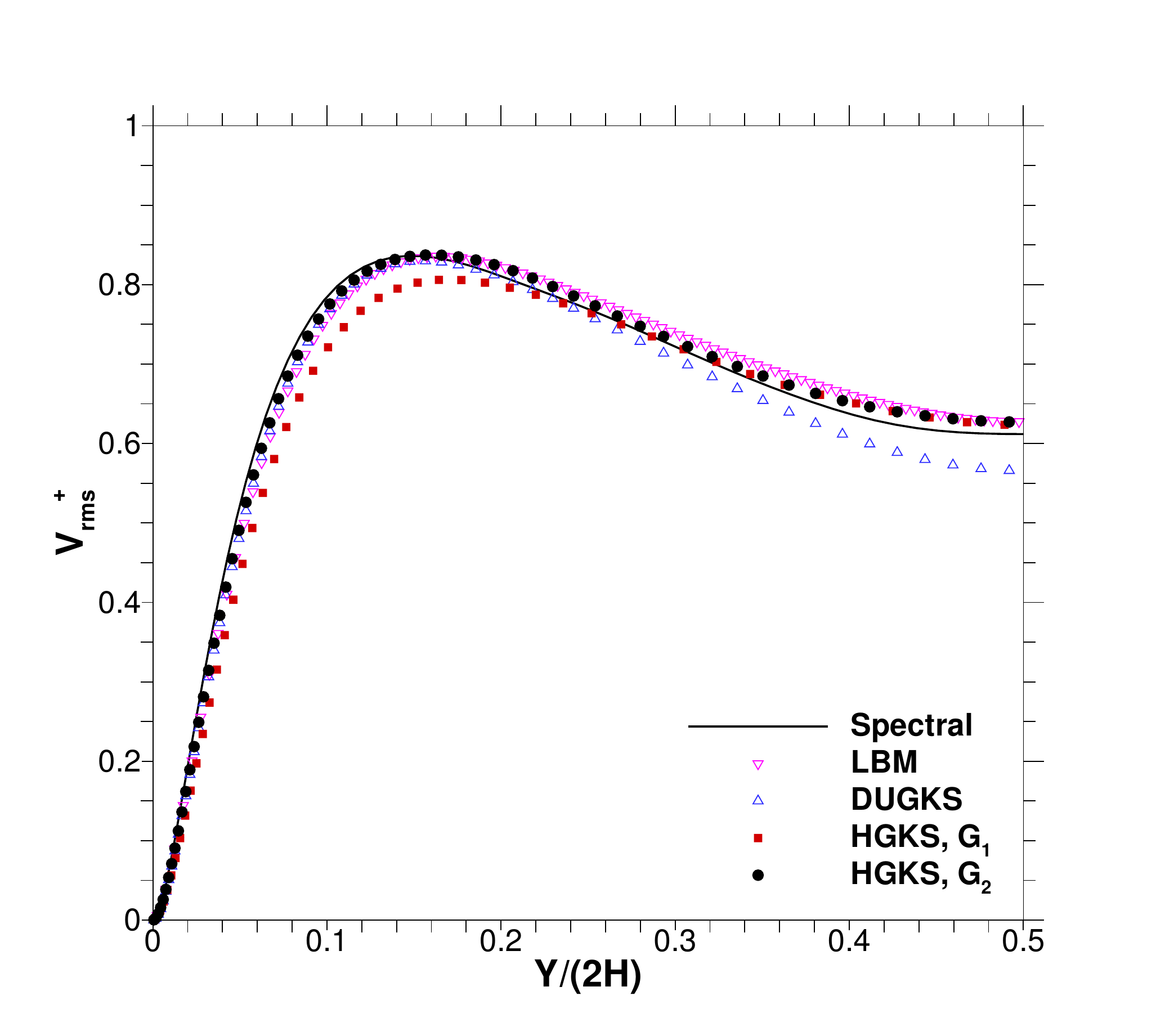}
	\includegraphics[width=0.45\textwidth]{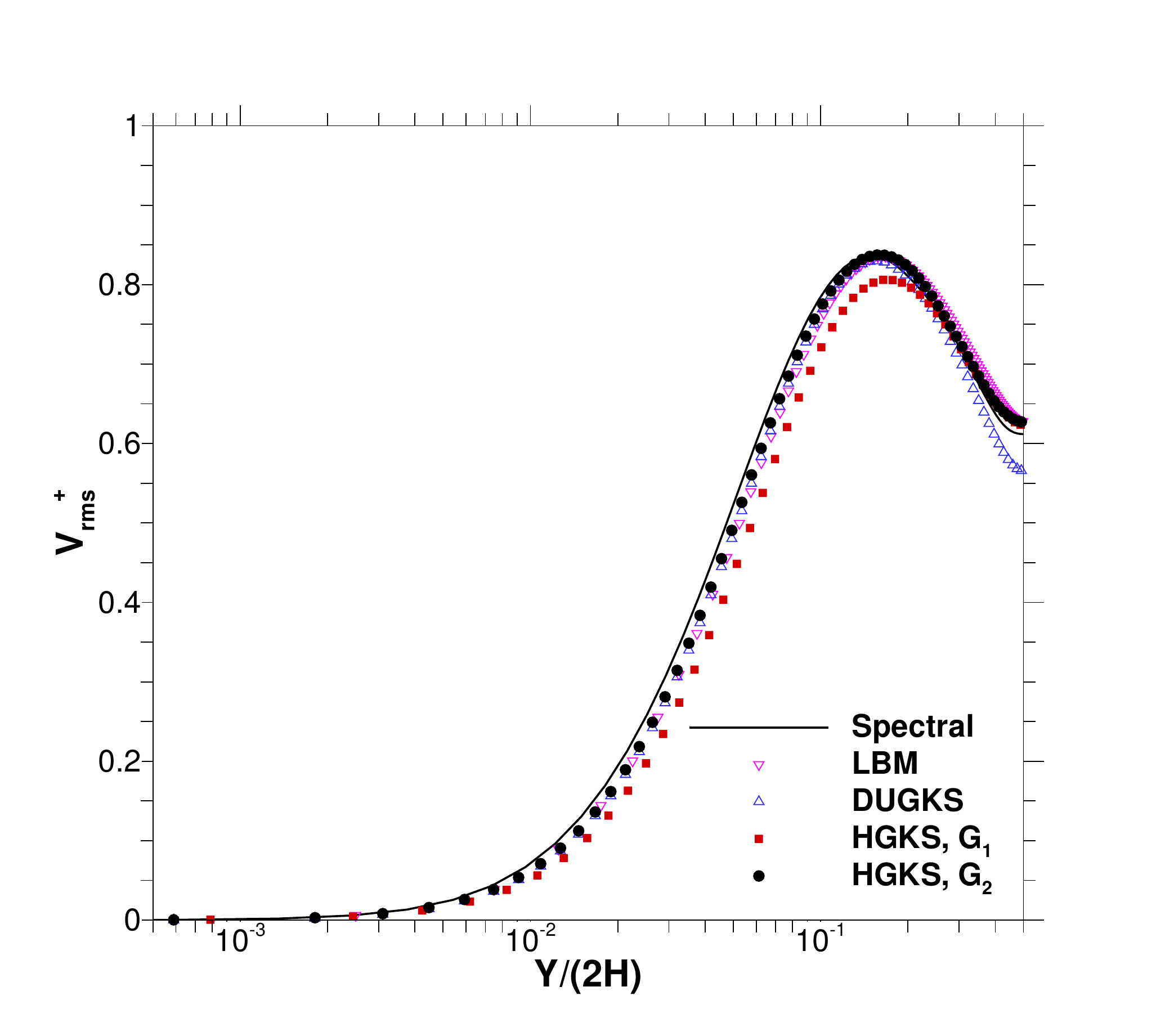}
	\includegraphics[width=0.45\textwidth]{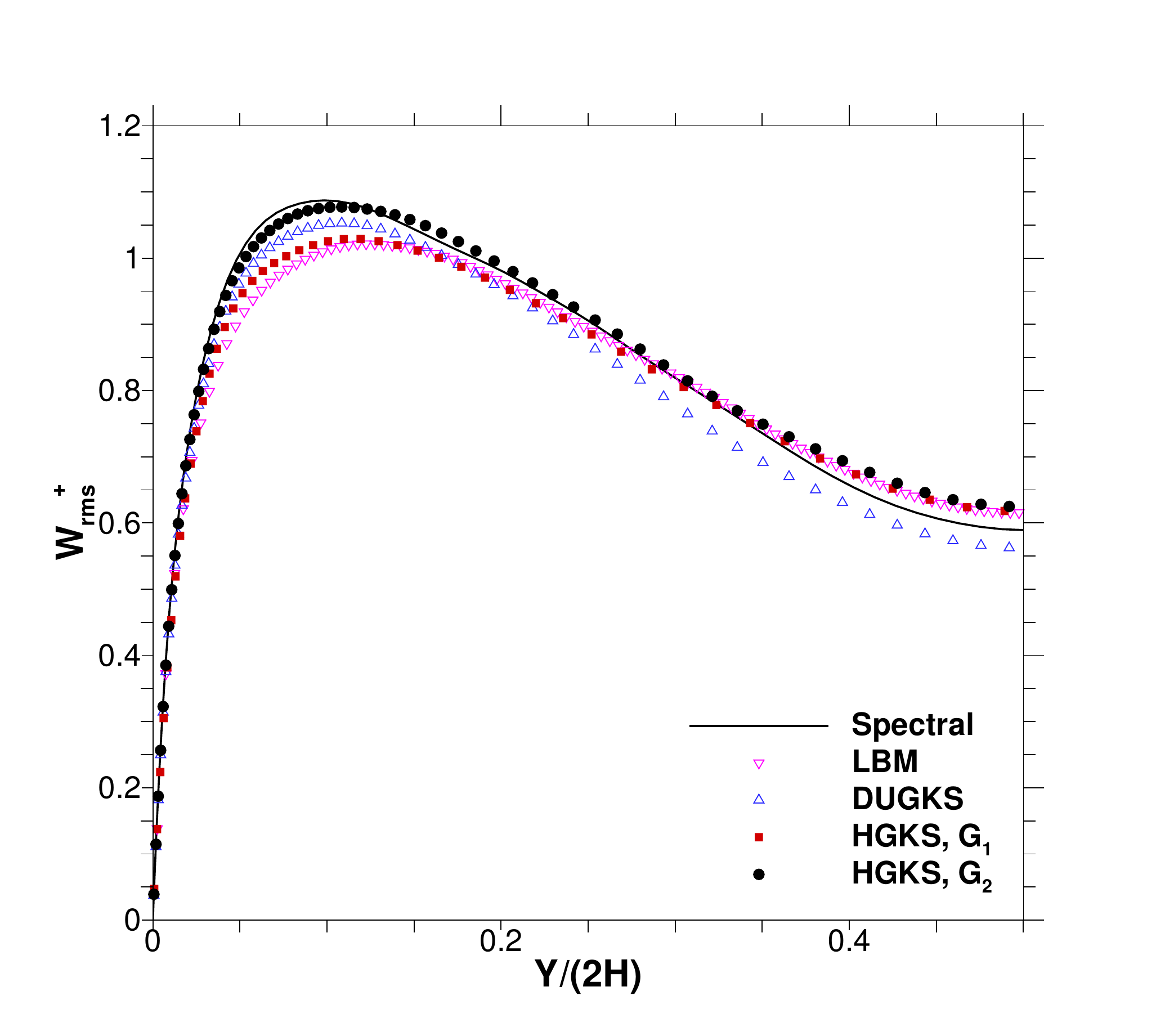}
	\includegraphics[width=0.45\textwidth]{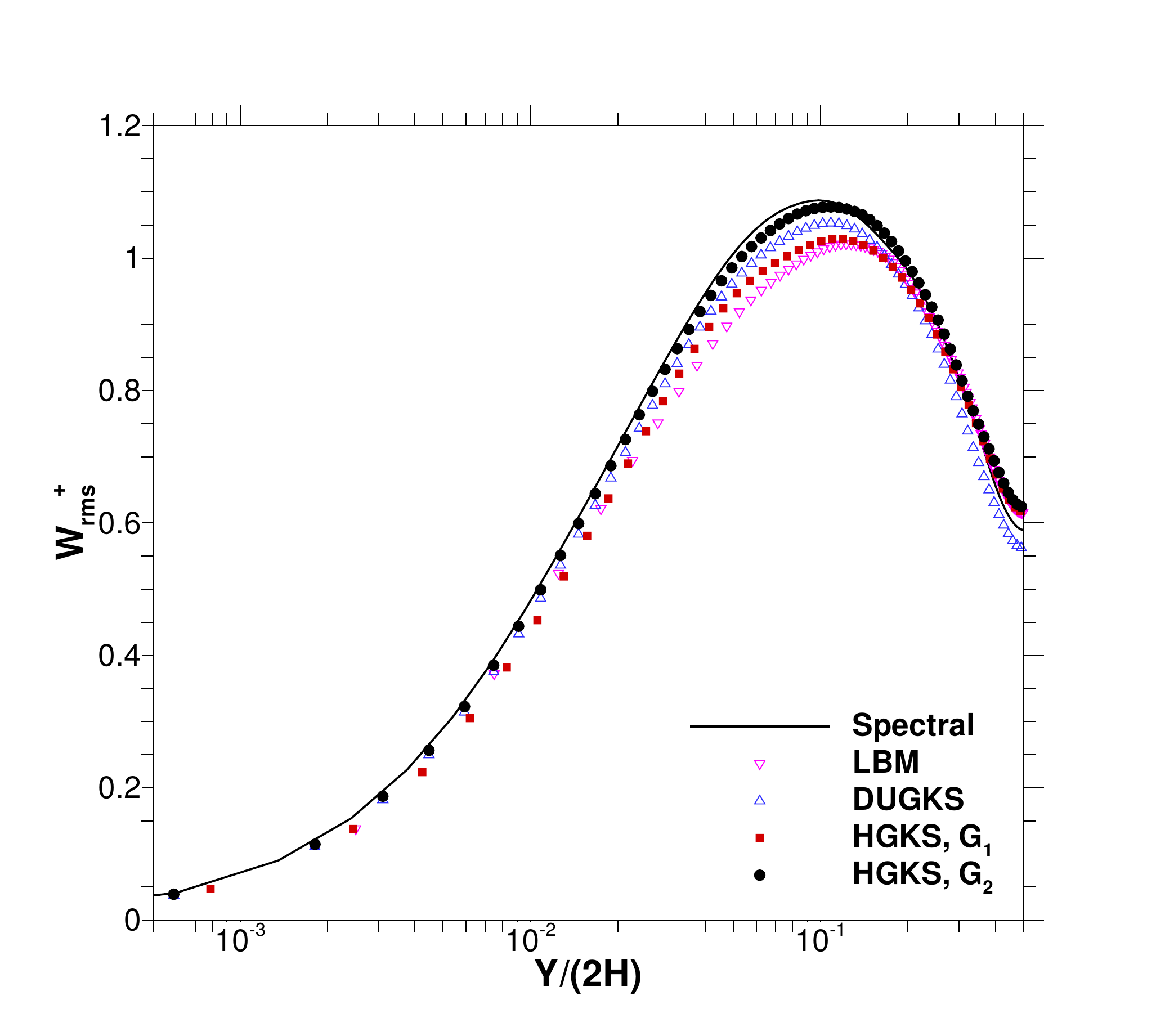}
	\caption{\label{channel_fluctuation} Turbulent channel flow: root-mean-square fluctuation velocity profiles in linear-linear and log-linear plots.}
\end{figure}
The averaged Reynolds shear stress profiles are shown in Fig.\ref{channel_velocity-2}
in linear-linear and log-linear plots.
The sum of Reynolds stress and viscous stress varies linearly from the channel center to the
channel wall. When compared to the spectral result, HGKS results
are clearly better than DUGKS and LBM results, especially in the near-wall region.
Turbulence intensities, i.e., the root-mean-square (rms) velocity
profiles are shown in Fig.\ref{channel_fluctuation} as well.
In the near-wall regions, the streamwise rms velocity is the largest and
the spanwise rms velocity is the smallest.
DUGKS yields a better result for $U_{rms}^+$ in the near-wall region.
Meanwhile, for $V_{rms}^+$ and $W_{rms}^+$, LBM behaves better in the near-center line region, and DUGKS behaves better in the near-wall region.
Considering the good agreement in near-wall region and the near-center line region with the spectral benchmark, it is clear that HGKS outweighs LBM and DUGKS.
The relative error in DUGKS is due only to the numerical truncation error, while the
relative error in LBM is due to both the domain size effect and
numerical truncation error \cite{DUGKS}. Different distribution of grid points in
the spanwise direction and different grid resolutions should be
tested, which could be a topic of HGKS for turbulent channel flows.
Finally, the limiting wall behavior of the Reynolds
stresses is shown in Table.\ref{table_grid_rms},
where $A_1 = U_{rms}^+/Y^+$, $B_1 = V_{rms}^+/Y^{+2}\times 10^2$,
 $C_1 = W_{rms}^+/Y^+$ and $D_1 = \left\langle -UV \right\rangle^+/Y^{+3}\times 10^3$. The $y$ behavior of tangential stresses $U_{rms}^+$ and $W_{rms}^+$
and the $y^2$ behavior of normal stress $V_{rms}^+$
are expected from consideration of no-slip boundary conditions
and continuity equation.
Reynolds shear stress $\left\langle -UV \right\rangle^+$ is in the asymptotic behavior of $y^3$,
which is regarded as the benchmark to calibrate the asymptotic near-wall behavior of eddy-viscosity turbulence models \cite{wilcox1998turbulence}.
Currently, the magnitudes of $A_1$, $B_1$, $C_1$ and $D_1$ are in good agreement with the results with spectral method \cite{kim1987turbulence}.

\begin{table}[!h]
    \centering
    \begin{tabular}{c|c|c|c|c}
        \hline \hline
        $Y^+$     &$A_1$     &$B_1$    &$C_1$  &$D_1$ \\
        \hline
        0.2124    & 0.3579   &0.8891   &0.1839 & 0.7827  \\
        \hline
        0.6505    & 0.3624   &0.7207   &0.1762 & 0.7046\\
        \hline
        1.1157    & 0.3628   &0.6568   &0.1678 & 0.7015\\
        \hline
        1.6096    & 0.3624   &0.6117   &0.1594 & 0.7065\\
        \hline
        2.1338    & 0.3614   &0.5716   &0.1512 & 0.7088\\
        \hline
        2.6902    & 0.3595   &0.5343   &0.1432 & 0.7055\\
        \hline
        3.2807    & 0.3564   &0.4990   &0.1353 & 0.6947\\
        \hline
        4.5715    & 0.3518   &0.4656   &0.1278 & 0.6756\\
        \hline
        5.2761    & 0.3453   &0.4339   &0.1205 & 0.6479\\
        \hline \hline
    \end{tabular}
    \caption{\label{table_grid_rms}
Turbulent channel flow: near-wall behavior of Reynolds
stresses for case $G_1$.}
\end{table}

In summary, the first simulation of turbulent channel flow using
HGKS shows that the results are reasonably accurate, and performance
is better than the second-order LBM and DUGKS. More importantly,
the HGKS results are obtained with a coarse grid resolution
covering a large domain size compared with LBM.
 As the equidistant grids are required for LBM, the grids displacement is limited to an extreme small value to resolve the viscous layer,
 i.e., $\Delta x^+ \approx 0.3$ in the whole computational domain, so the required grids number of LBM will exceed HGKS dramatically.
In the current computation, the
spatial reconstruction is based on the WENO reconstruction on
uniform meshes due to the small variation of neighboring cells, and
the order of accuracy may be affected slightly. Currently, we are
working on the genuinely high-order gas-kinetic scheme on the
nonuniform and curvilinear meshes, and the geometrical errors can be
excluded in the future simulations.

\subsection{Isotropic compressible turbulence}
The isotropic compressible turbulence is regarded as one of
cornerstones to elucidate the effects of compressibility for
turbulence \cite{sarkar1991analysis, samtaney2001direct}. Based on the
numerical experiments and theoretical analysis, the isotropic
compressible turbulence is divided into four main dynamical regimes,
i.e. the low-Mach number quasi-isentropic regime, the low-Mach
number thermal regime, the nonlinear subsonic regime, and the
supersonic regime \cite{sagaut2008homogeneous}.
High-order compact finite difference method \cite{lele1992compact} has been widely utilized in the simulation of
isotropic compressible turbulence with moderate turbulent Mach
number, i.e. $Ma_t \leq 0.8$. However, when simulating the turbulent
in supersonic regime, the compact scheme fails to capture strong shocklets and
suffers from numerical instability.
In this case, we concentrate on
the decaying isotropic compressible turbulence without external
force. The flow domain of numerical simulation is a cube box $0 \leq
x, y, z\leq 2 \pi$, with periodic boundary conditions in all three
Cartesian directions for all flow variables. A three-dimensional
solenoidal random initial velocity field $\bm{U}$ is generated by a
specified spectrum \cite{passot1987numerical}
\begin{align}\label{initial_spectrum}
E(\kappa) = A_0 \kappa^4 \exp (-2\kappa^2/\kappa_0^2),
\end{align}
where $A_0$ is a constant to get a specified initial kinetic energy,
$\kappa$ is the wave number, $\kappa_0$ is the wave number at which
the spectrum peaks. In this paper, fixed $A_0$ and $\kappa_0$  in
Eq.(\ref{initial_spectrum}) are chosen for all cases, which are
initialized by $A_0 = 0.00013$ and $ \kappa_0 = 8$.
Evolution of this artificial system is determined by
initial thermodynamic quantities and two dimensionless parameters,
i.e. the initial Taylor microscale Reynolds number and turbulent Mach number
\begin{align*}
Re_{\lambda}&=\frac{(2 \pi)^{1/4}}{4} \frac{\rho_0}{\mu_0}\sqrt{2
    A_0}\kappa_0^{3/2},\\
&Ma_t=\frac{\sqrt{3}}{\sqrt{\gamma R T_0}}u_{rms},
\end{align*}
where the initial density $\rho_0 = 1$ and $U_{rms}$ is the root
mean square of initial velocity field
\begin{align*}
U_{rms}=\left\langle \frac{\bm{U} \cdot \bm{U}}{3}
\right\rangle^{1/2}.
\end{align*}
With $Re_{\lambda}$, $Ma_{t}$ and $\gamma = 1.4$, the initial
viscosity $\mu_0$, pressure $p_0$ and temperature  $T_0$ can be
determined. The dynamic velocity can be also given by
\begin{align*}
\mu=\mu_0(\frac{T}{T_0})^{0.76}.
\end{align*}
With current initial strategy, the initial ensemble turbulent
kinetic energy $K_0$, ensemble enstrophy $\Omega_0$, ensemble
dissipation rate $\varepsilon_0$, large-eddy-turnover time
$\tau_{to}$, Kolmogorov length scale $\eta_0$, and the Kolmogorov
time scale $\tau_0$ are given as
\begin{equation}\label{initial_def}
\begin{aligned}
K_0=&\frac{3A_0}{64} \sqrt{2 \pi} \kappa_0^5, ~ \Omega_0=\frac{15
    A_0}{256} \sqrt{2 \pi} \kappa_0^7,~
\tau_{to}=\sqrt{\frac{32}{A_0}}(2 \pi)^{1/4} \kappa_0^{-7/2},\\
&\varepsilon_0=2\frac{\mu_0}{\rho_0} \Omega_0, ~
\eta_0=(\nu_0^3/\varepsilon_0)^{1/4}, ~
\tau_0=(\nu_0/\varepsilon_0)^{1/2}.
\end{aligned}
\end{equation}
For the compressible isotropic turbulence, starting from the initial
flows, the large eddies transfer their turbulent kinetic energy
successively to smaller eddies.
The time history of the root-mean-square density fluctuation $\rho_{rms}(t)$, turbulent kinetic energy $K(t)$,
skewness factor $S_u(t)$ and flatness factor $F_u(t)$ for
velocity slope are defined as
\begin{align*}
\rho_{rms}(t)&=\sqrt{\left\langle(\rho-\overline{\rho})^2\right\rangle},\\
K(t)&=\frac{1}{2}\left\langle\rho \boldsymbol{U}\cdot \boldsymbol{U}\right\rangle,\\
S_u(t)=&\sum_i\frac{\left\langle(\partial_i U_i)^3\right\rangle}{\left\langle(\partial_i
    U_i)^2\right\rangle^{3/2}},\\
F_u(t)=&\sum_i\frac{\left\langle(\partial_i U_i)^4\right\rangle}{\left\langle(\partial_i
    U_i)^2\right\rangle^{2}}.
\end{align*}
In this process, the evolution of
turbulent kinetic energy is of interest since it is a fundamental
benchmark for incompressible and compressible turbulence modeling
\cite{yoshizawa1985statistically, pope2001turbulent}.
The decay of the ensemble
turbulent kinetic energy can be described approximately by
\cite{sarkar1991analysis}
\begin{equation}\label{dkdt}
\begin{aligned}
\frac{\text{d}\left\langle K\right\rangle}{\text{d}t}=\varepsilon&+\left\langle p \theta \right\rangle,\\
\varepsilon=\varepsilon_s+&\varepsilon_d,
\end{aligned}
\end{equation}
where $\varepsilon_s=\left\langle \mu \omega_i
\omega_i\right\rangle$ is the ensemble solenoidal dissipation rate,
$\displaystyle\varepsilon_d= 4/3\left\langle\mu
\theta^2\right\rangle$ is the ensemble dilational dissipation rate,
$\left\langle p \theta \right\rangle$ is the ensemble
pressure-dilation transfer,
$\displaystyle\omega_i=\epsilon_{ijk}\partial U_k/\partial x_j$ is
the fluctuating vorticity, $\epsilon_{ijk}$ is the alternating
tensor and $\theta = \nabla \cdot \bm{U}$ is the fluctuating
divergence of velocity.

\begin{table}[!h]
    \centering
    \begin{tabular}{c|c|c|c|c|c}
        \hline \hline
            Case      &$Ma_{t}$   &$\text{d}t_{ini}/\tau_{to}$  &$\kappa_{max}\eta_0$   &$\Delta/\eta_0$ &$\Delta/\lambda_0$\\
         \hline
            $R_1$     &0.8        &1.04/1000   &3.613 &0.819   &2.551\\
         \hline
            $R_2$     &1.0        &1.09/1000   &3.613 &0.819   &2.041\\
         \hline
            $R_3$     &1.2        &1.14/1000   &3.613 &0.819   &1.700\\
         \hline
            $R_4$     &1.6        &1.19/1000   &3.613 &0.819   &1.275\\
        \hline \hline
    \end{tabular}
 \caption{\label{re72gridtable} Isotropic compressible turbulence: parameters for different high turbulent Mach number.}
\end{table}

\begin{figure}[!h]
\centering
\includegraphics[width=0.475\textwidth]{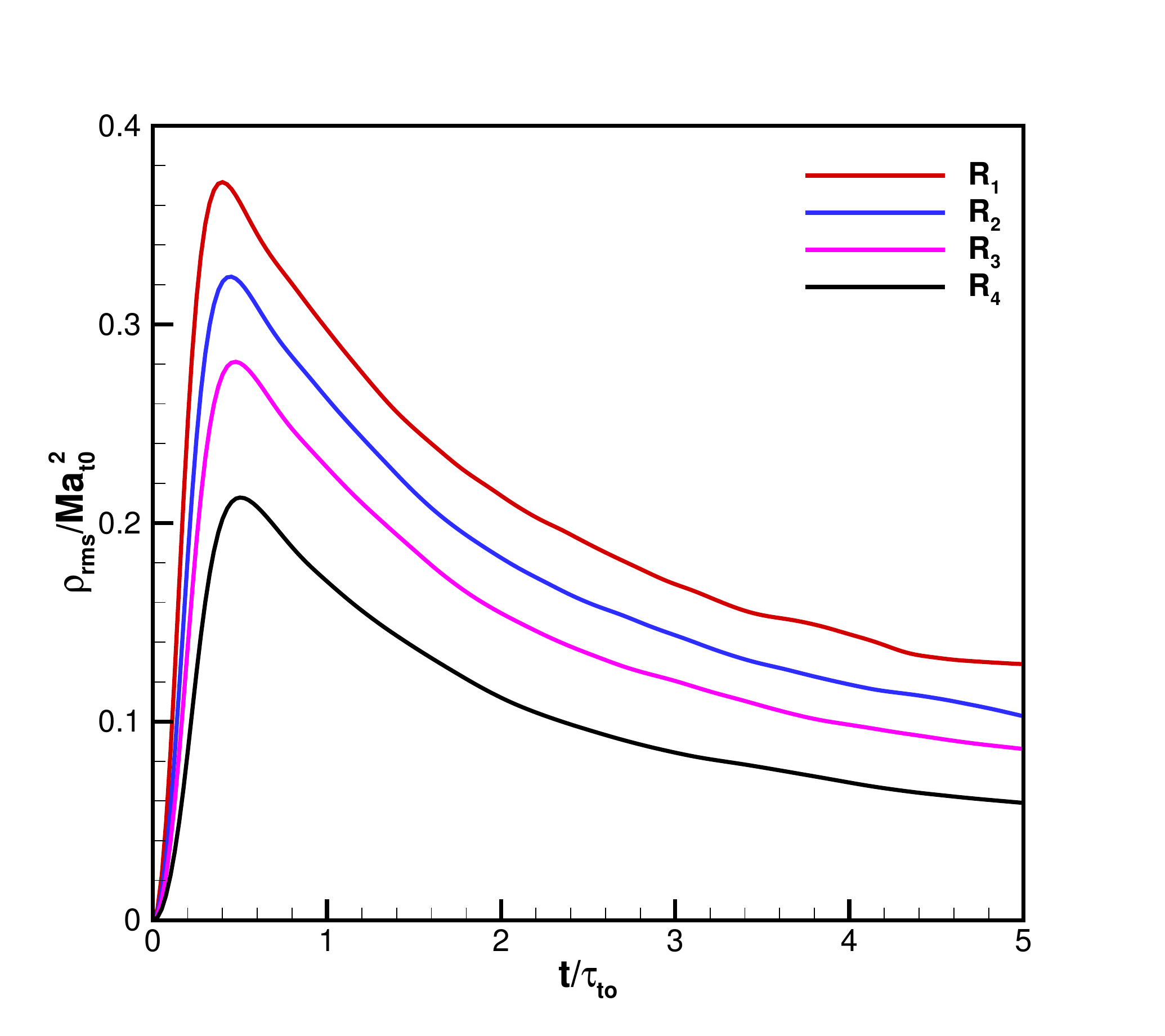}
\includegraphics[width=0.475\textwidth]{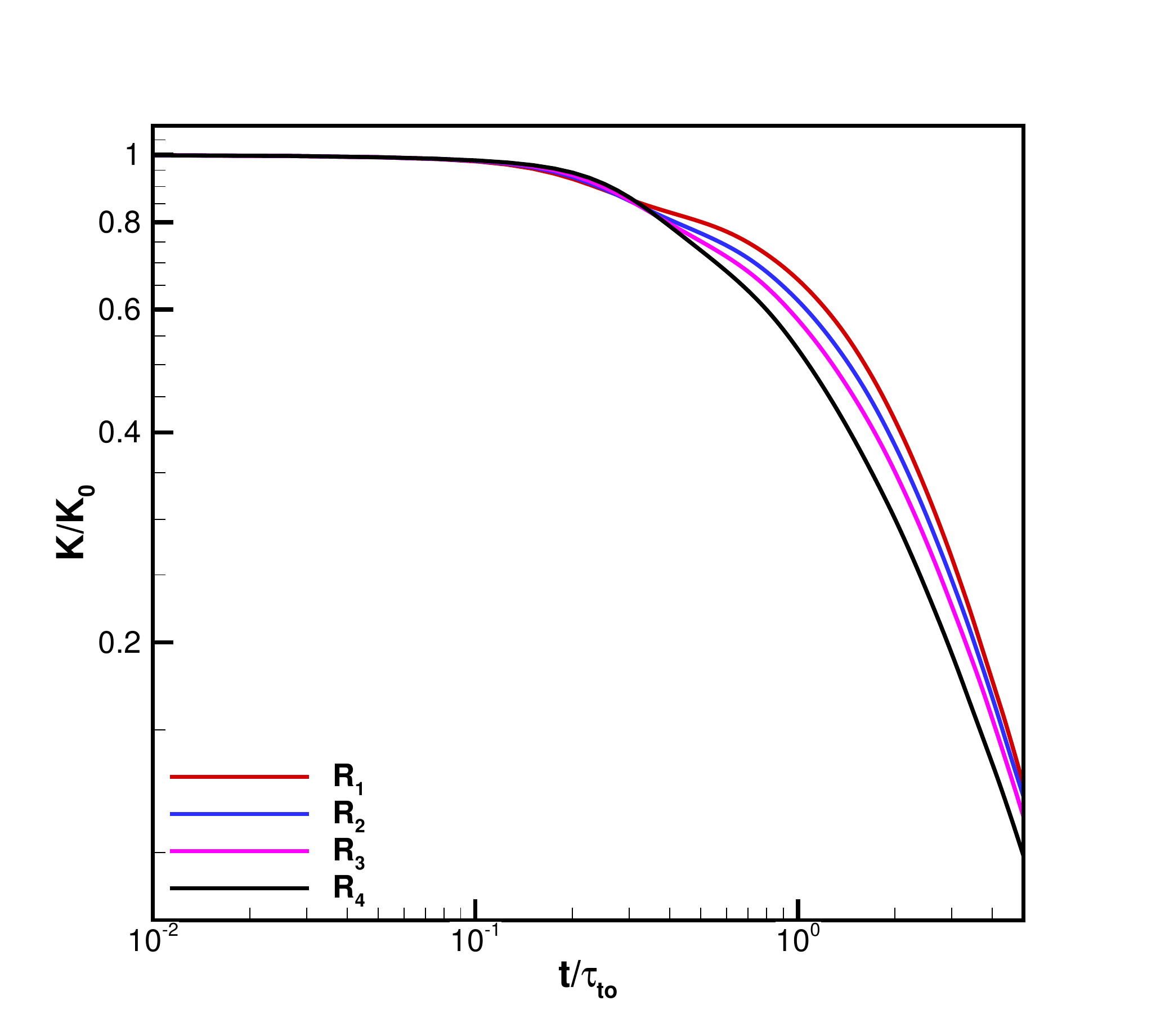}\\
\includegraphics[width=0.475\textwidth]{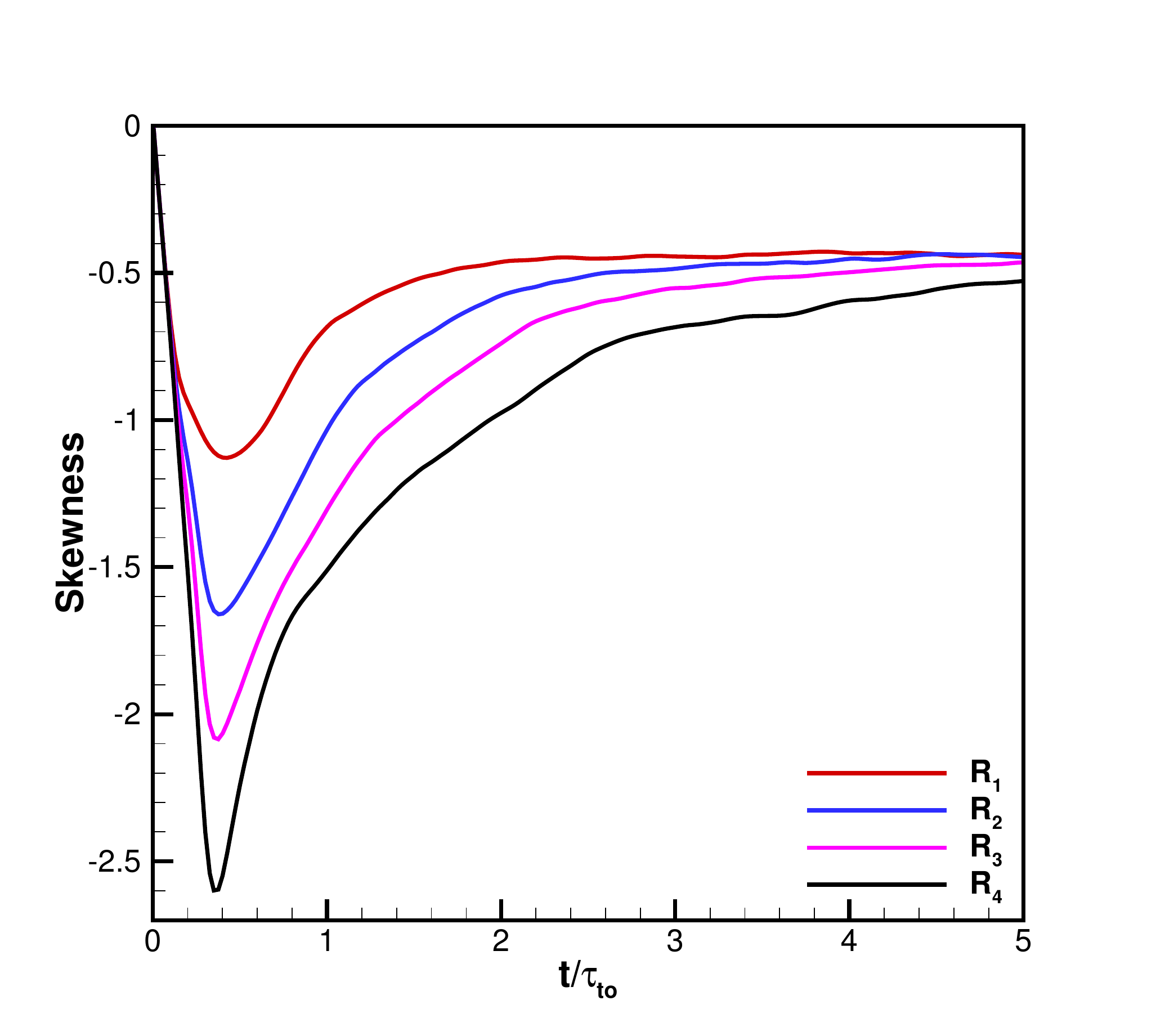}
\includegraphics[width=0.475\textwidth]{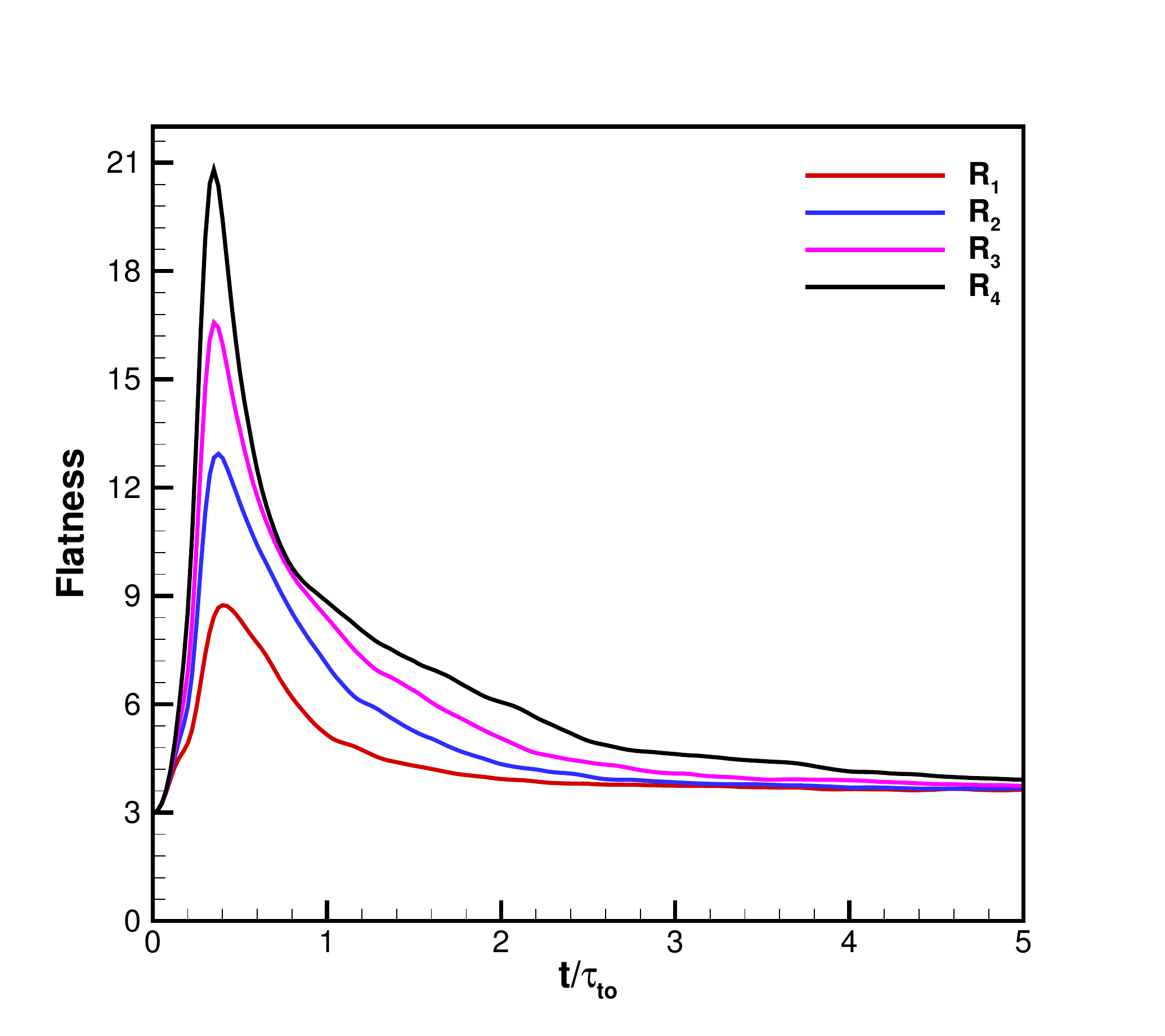}
\caption{\label{history_ma08_12_k} Isotropic compressible turbulence: time history of
$\rho_{rms}/Ma_{t}^2$, $K/K_0$, $S_u$, and $F_u$ for cases $R_1$-$R_4$.}
\end{figure}

Due to the robustness of current scheme, we can simulate this case
up to the supersonic regime, which is seldom reported in literatures.
The WENO-Z scheme \cite{WENO-Z} used for the spatial reconstruction, no extra
special treatment is needed in the code. In this case, the numerical
tests $R_1$, $R_2$, $R_3$ and $R_4$ are presented with a fixed
Taylor micro-scale Reynolds number $Re_{\lambda} = 72$ to and the turbulent Mach number
form $Ma_t = 0.8$ to $Ma_t = 1.6$.
In the computation,   $256$ cores are used for $512^3$ uniform cells,
and more parameters are given in Table.\ref{re72gridtable},
where $\lambda_0$ is the initial mean free path
approximated by $\mu_0 = 1/3 \rho_0 c_0 \lambda_0$
\cite{GKS-Xu3}, $\Delta$ is the uniform grid size in each
direction, $\eta_0$ is the initial Kolmogorov length scale as in
Eq.(\ref{initial_def}), $\kappa_{max} = \sqrt{2} \kappa_0 N/3$ is
the maximum resolved number wave number, $\kappa_0 = 8$ as
Eq.(\ref{initial_spectrum}) and $N$ is the number of grid points in
each Cartesian direction.
The numerical tests show that the minimum spatial resolution parameter
$\kappa_{max} \eta_0 \ge 2.71$ and the maximum temporal resolution
parameter $\Delta t_{ini}/\tau_{to} \leq 5.58/1000$ for HGKS is
adequate for resolving the isotropic compressible turbulence \cite{GKS-high-4}.
 According to Table.\ref{re72gridtable},
the Kolmogorov length scale is still larger than the
mean free path, and each grid always contains more than one mean free path.
This provides the intuitive evidence for controversial issue that smallest eddies in
turbulence may still within the framework of continuum mechanics
assumption.

\begin{figure}[!h]
    \centering
\includegraphics[width=0.475\textwidth]{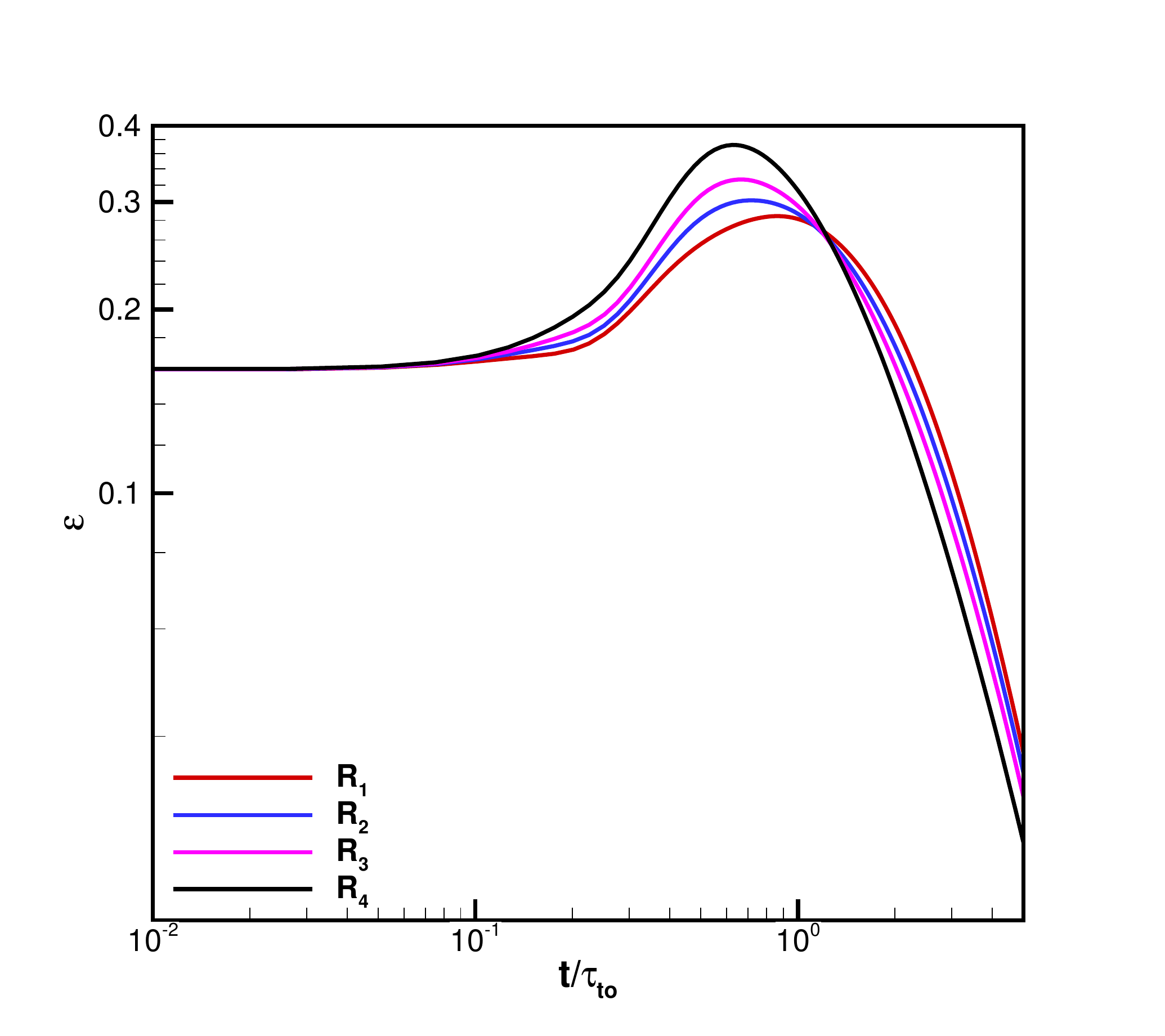}
\includegraphics[width=0.475\textwidth]{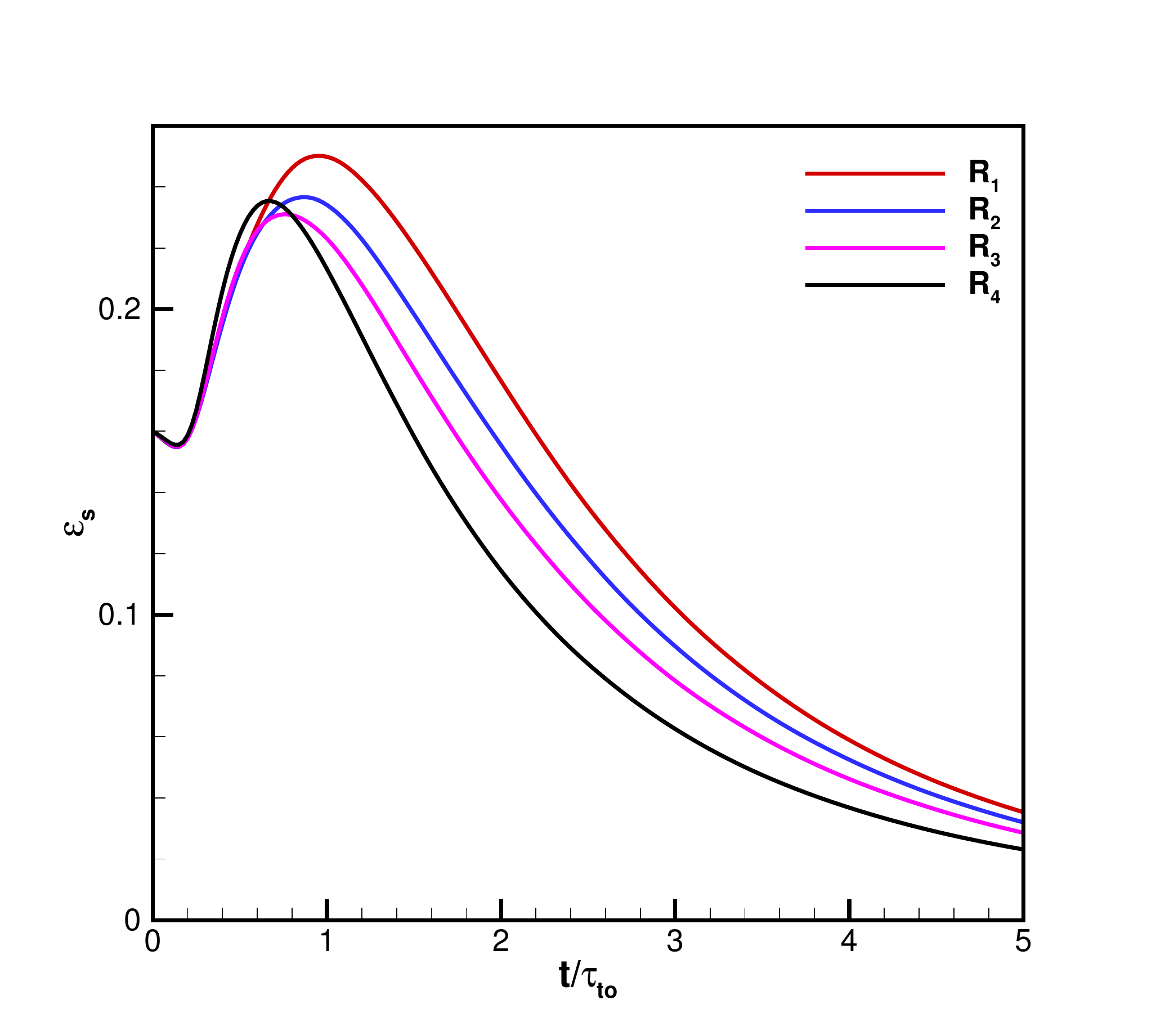}\\
    \includegraphics[width=0.475\textwidth]{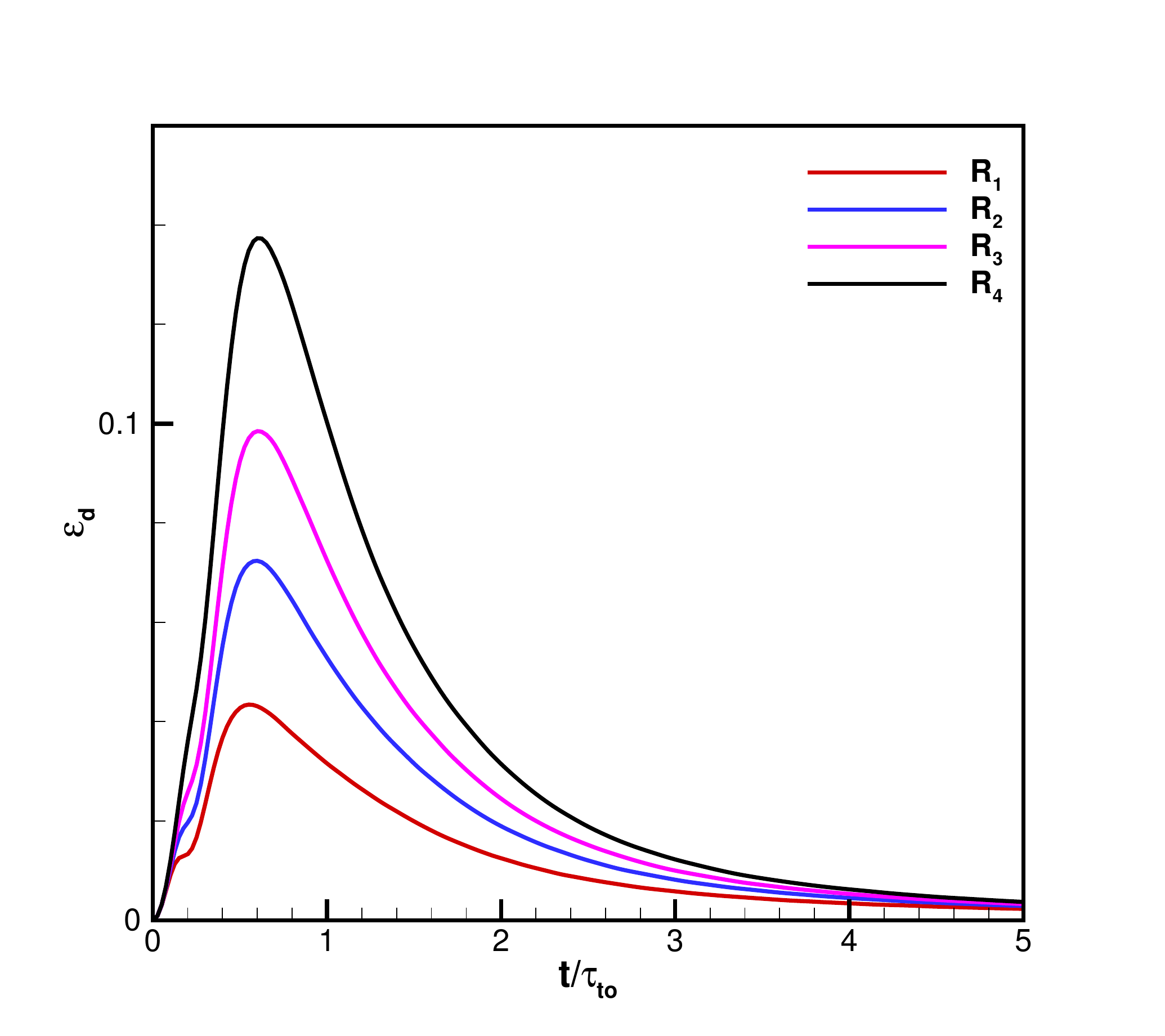}
    \includegraphics[width=0.475\textwidth]{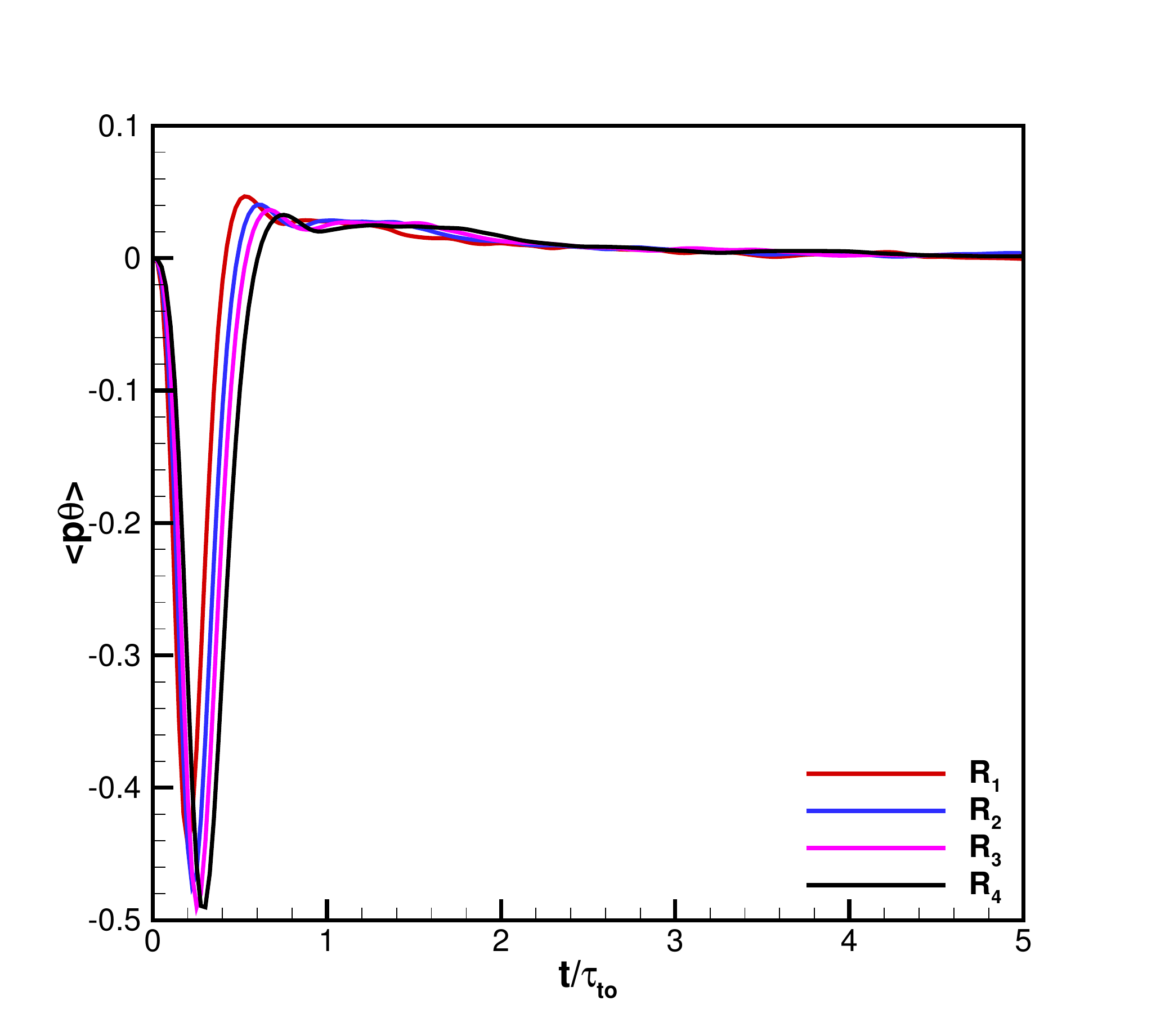}
    \caption{\label{history_ma08_12} Isotropic compressible turbulence: time history of $\varepsilon$, $\varepsilon_s$, $\varepsilon_d$ and $\left\langle p \theta \right\rangle$ for cases
        $R_1$-$R_4$.}
\end{figure}

\begin{figure}[!h]
\centering
\includegraphics[width=0.475\textwidth]{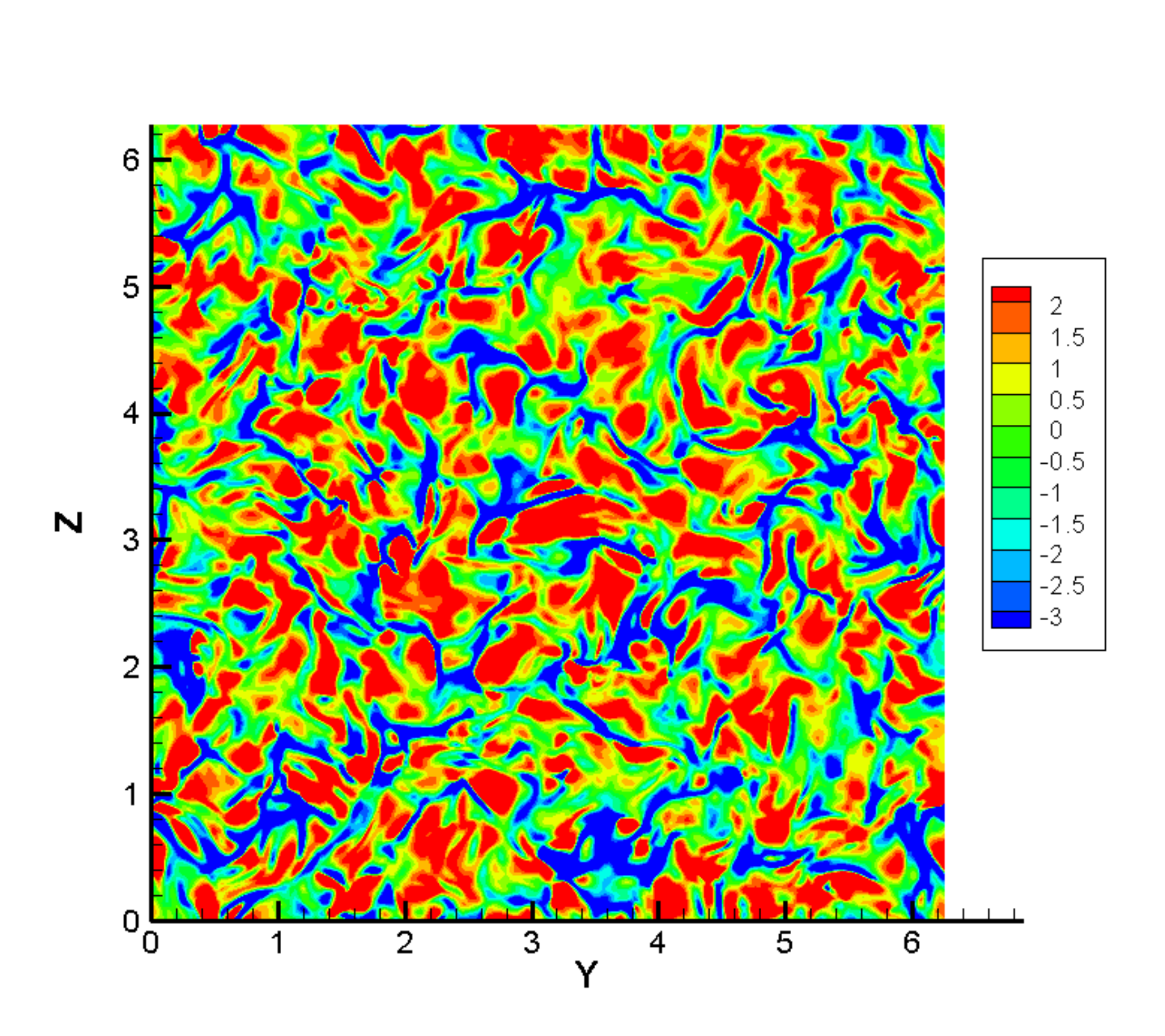}
\includegraphics[width=0.475\textwidth]{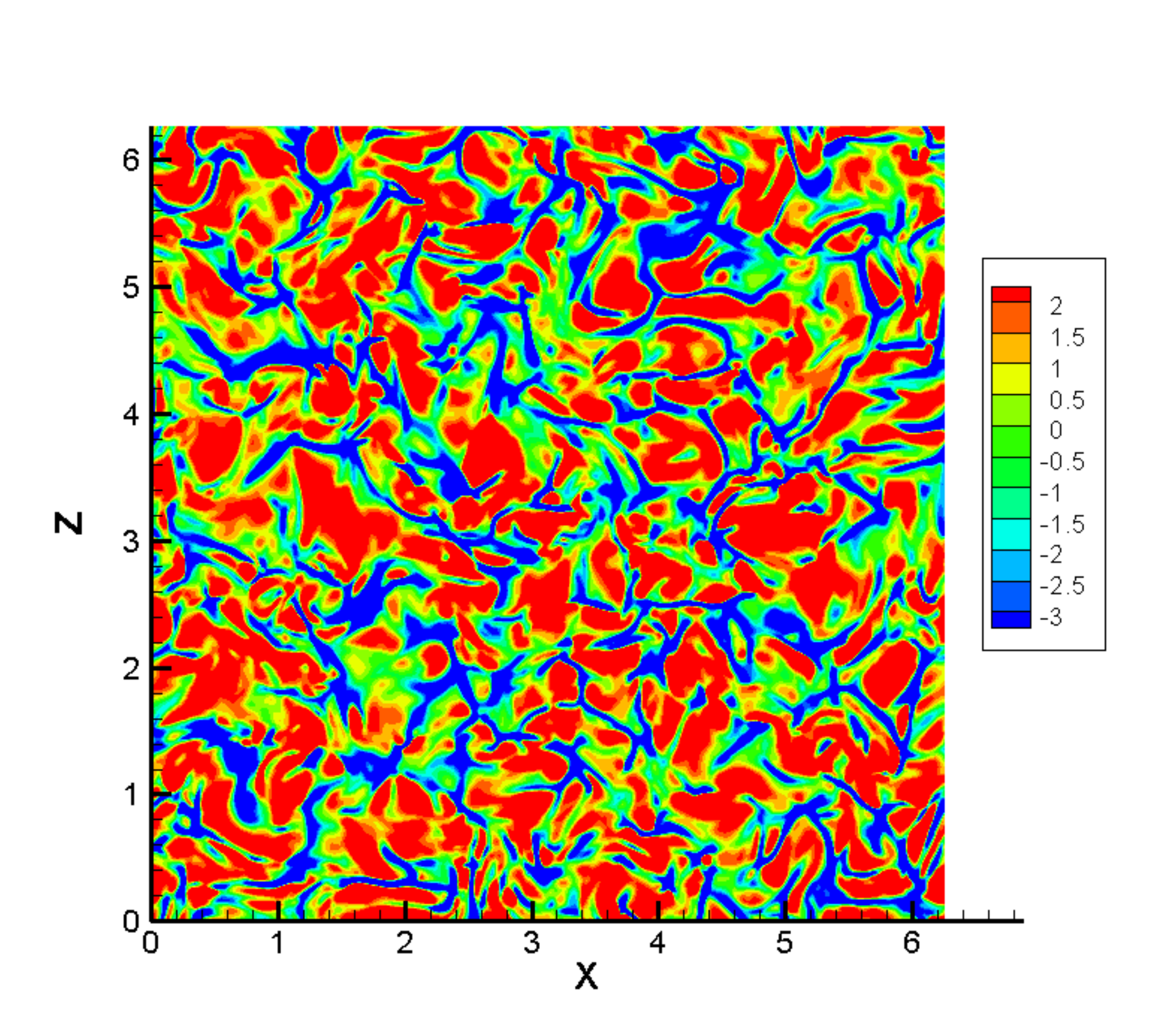}\\
\includegraphics[width=0.475\textwidth]{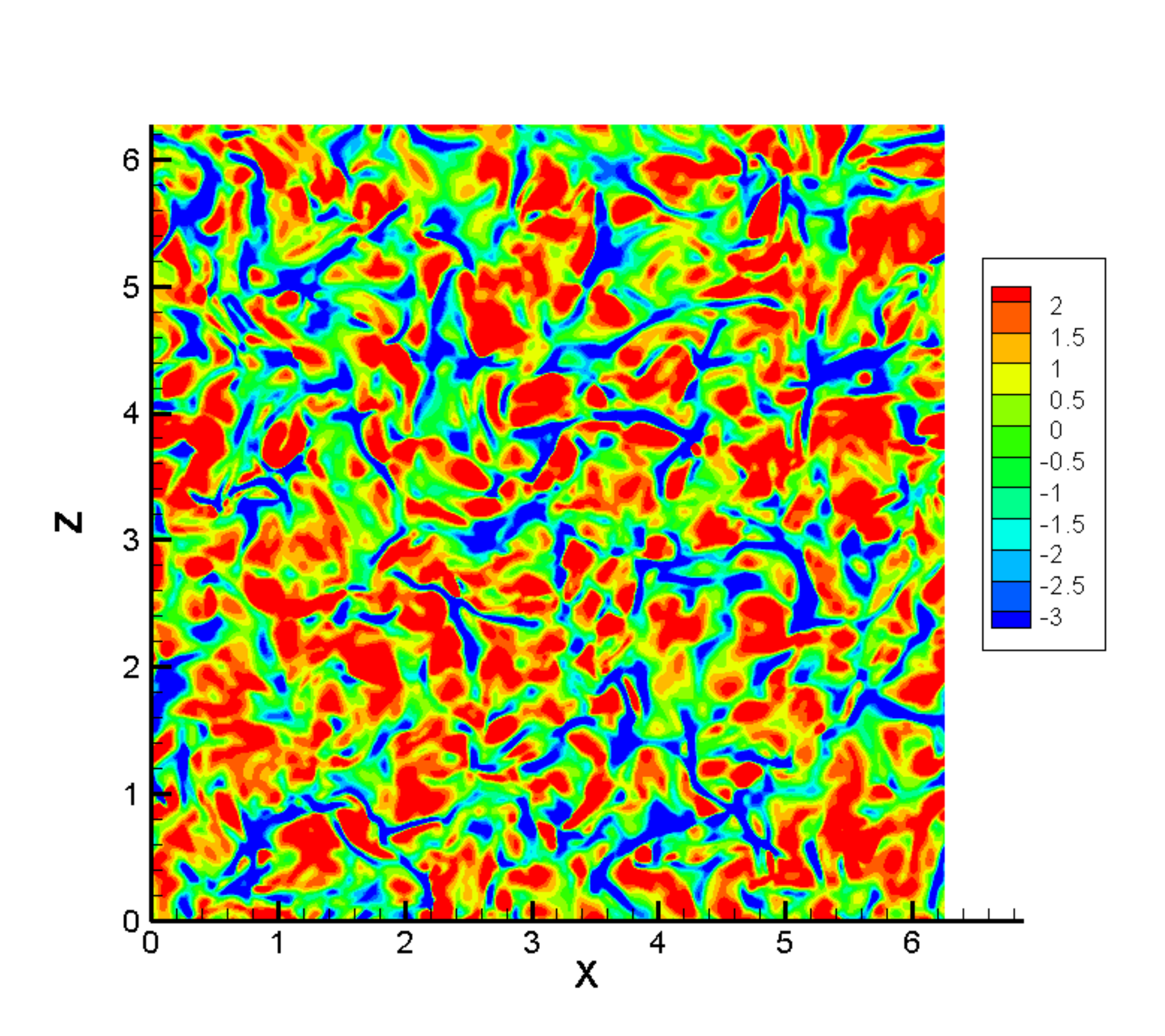}
\includegraphics[width=0.475\textwidth]{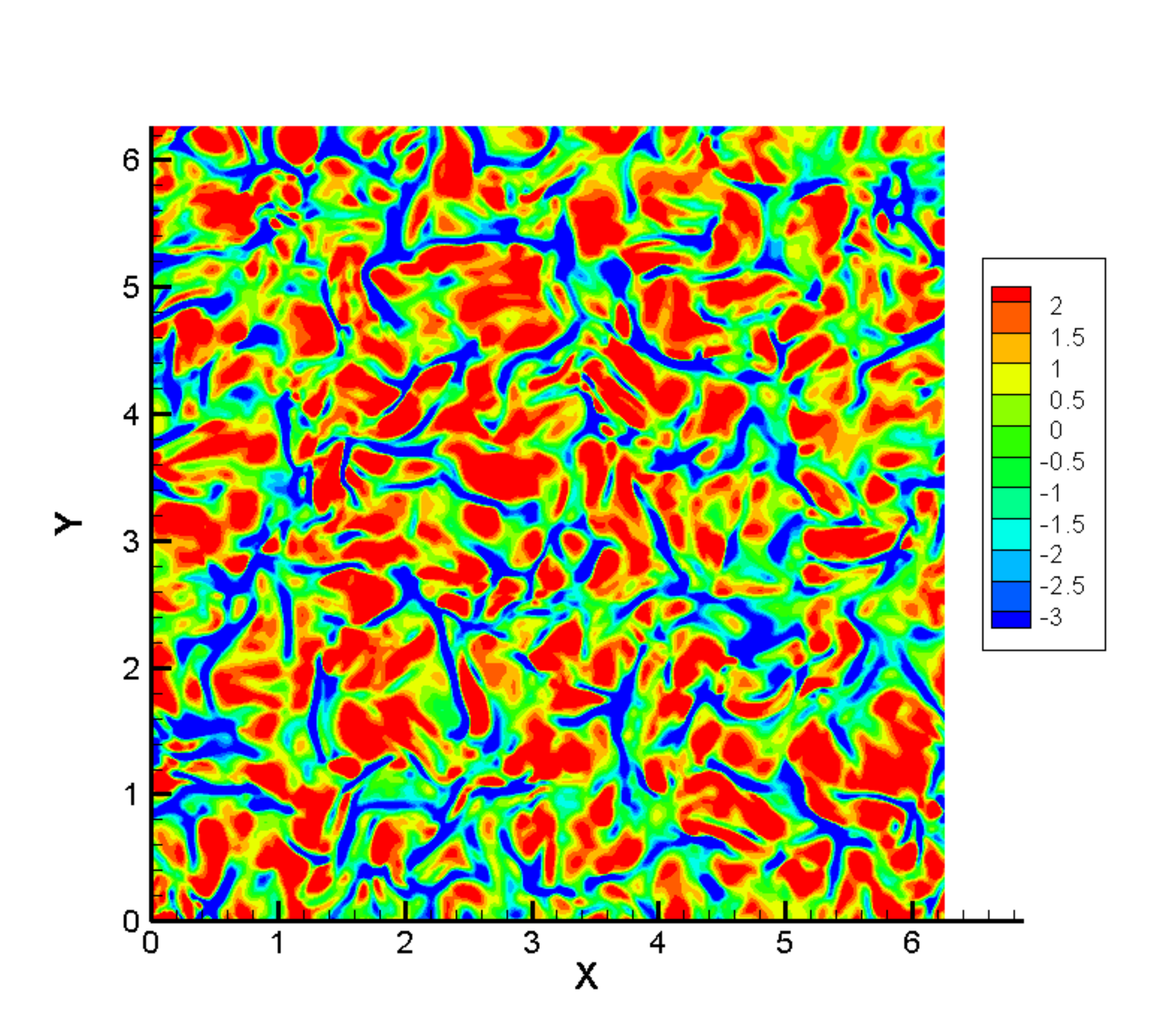}
\caption{\label{Ma08theta} Isotropic compressible turbulence: contours of normalized dilation
    $\theta/\left\langle \theta \right\rangle^{\ast}$ for cases $R_1$-$R_2$ at $t/\tau_{to} = 1.0$, and cases $R_3$-$R_4$ at $t/\tau_{to} = 2.0$.}
\end{figure}

Statistical quantities are provided for these cases, which provide
benchmark solutions for supersonic isotropic compressible
turbulence. Time history of  $\rho_{rms}(t)/Ma_{t}^2$, $K(t)/K_0$, $S_u(t)$
and $F_u(t)$ are presented in Fig.\ref{history_ma08_12}, which
provides benchmark data for simulating isotropic compressible
turbulence up to supersonic regime.
The normalized root-mean-square
density $\rho_{rms}/Ma_{t}^2$ decreases monotonically with the
increase of initial turbulent Mach number.
With the increase of turbulent Mach number, the peak of skewness and
flatness factor deviate from those of the low-Mach number thermal regime ($Ma_t \leq 0.3$) severely.
These large deviation indicates the most significant flow structures of
isotropic compressible turbulence resulting from the shocklets.
As the initial turbulent
Mach number increases, the peak of dissipation increases as well.
Obviously, ensemble solenoidal dissipation rate $\varepsilon_s$
decreases with the increase of $Ma_{t}$, while the dilational
dissipation rate $\varepsilon_d$ rises with the increase of $Ma_{t}$.
In addition, $\left\langle p \theta \right\rangle$ changes
signs during the evolution and preserves small but positive
value thereafter, which agree with earlier study for subsonic isotropic
turbulence \cite{sarkar1991analysis}.
During the early stage of the decaying
supersonic isotropic turbulence, the ensemble pressure-dilation term
can be in the same order of ensemble total dissipation rate.
It is reported that the ratio between the ensemble pressure-dilation
term and the right hand side of Eq.(\ref{dkdt}) becomes small for
solenoidal forced quasi-stationary supersonic isotropic turbulence
\cite{wang2018kinetic}.

To investigate the behavior of supersonic isotropic compressible
turbulence, the contours of normalized dilation $\theta/\left\langle
\theta \right\rangle^{\ast}$ on $x=0/y=0/z=0$
slices are presented in Fig.\ref{Ma08theta} for four
cases, where $\left\langle \theta
\right\rangle^{\ast}$ is root-mean-square dilation.
Contours of normalized dilation show very different behavior
between the compression motion and expansion motion. Strong
compression regions $\theta/\left\langle \theta\right\rangle^{\ast}
\leq -3$ are usually recognized as shocklets
\cite{samtaney2001direct}.  In current study, shocklets behave in
the shape of narrow and long ribbon, while high expansion regions
$\theta/\left\langle \theta \right\rangle^{\ast} \geq 2$ are in the
type of localized block. In addition, strong compression regions are
close to several regions of high expansion. This behavior is
consistent with the physical intuitive that expansion regions can be
identified just downstream of shock waves \cite{wang2018effect}.
These random distributed shocklets and high expansion region lead to
strong spatial gradient in flow fields, which pose much greater
challenge for high-order schemes when implementing DNS for isotropic
turbulence in supersonic regime.   Numerically, few methods survive from such tough cases.
The isotropic compressible turbulence with high turbulent Mach number up to supersonic regime has been studied, which
verifies that HGKS provides a valid tool for numerical and physical studies of compressible turbulence in
supersonic regime. More challenging compressible turbulence problems will be investigated in the future, such as shock-boundary
interaction and supersonic turbulent boundary layer.

\section{Conclusion}
Based on the multi-scale physical transport and the coupled temporal-spatial gas evolution,
the HGKS provides a useful tool for the numerical study of compressible turbulent flow. 
The performance of HGKS has been fully investigated for the DNS
of isotropic compressible turbulence up to the supersonic regime.
In order to increase the scale of computation, a parallel code of HGKS has been constructed with
domain decomposition and MPI implementation.  
The resulting scheme is tested
for Taylor-Green vortex problem, turbulent channel flow and
isotropic compressible turbulence.
It is the first successful DNS application of HGKS for turbulent flow from nearly incompressible to supersonic one. 
The scalability of parallel computation is
validated, and the computational cost is comparable with the
high-order finite difference method. For the nearly incompressible
turbulent flow, the performance of HGKS is also comparable with the
finite difference method. 
Based on the accuracy of the numerical solution, the numerical dissipation of the scheme in the turbulence simulation is quantitatively evaluated.
As a mesoscopic method, the HGKS performs
better than both LBM and DUGKS. 
More importantly, HGKS shows special advantages for the supersonic turbulence due to the accuracy and
robustness. More challenging examples using HGKS at higher Reynolds
numbers and different flow configurations will be investigated
in the future.

\section*{Ackonwledgement}
This research is supported by National Natural Science Foundation of
China (11701038, 11772281, 91852114), the Fundamental Research
Funds for the Central Universities, and the National Numerical Windtunnel project. The authors would like to thank
Prof. Xuesheng Chu for implementation of parallel
computation, Prof. Lianping Wang for providing the channel turbulence data of LBM and DUGKS,
and TianHe-II in Guangzhou for providing high
performance computational resources.

\end{document}